\documentclass[aps,prb,10pt,twocolumn,floatfix,showpacs,superscriptaddress]{revtex4-1}
\usepackage[colorlinks=true,citecolor=blue,linkcolor=blue,urlcolor=blue]{hyperref}
\usepackage{amsmath}
\usepackage{amssymb}
\usepackage{graphicx,amsmath,amsfonts,bm}
\usepackage{epstopdf}
\usepackage{bm}
\usepackage{bbm}
\usepackage{color}
\usepackage{relsize}
\usepackage{dsfont}
\usepackage{braket}
\usepackage[caption=false]{subfig}
\usepackage{hyperref}
\usepackage[all]{hypcap} 
\usepackage[T1]{fontenc}
\usepackage{dsfont}
\usepackage{mathbbol}

\begin{document}
\title{Localized states coupled to a network of chiral modes in minimally twisted bilayer graphene} 
\author{P. Wittig}
\affiliation{Institute for Mathematical Physics, TU Braunschweig, 38106 Braunschweig, Germany} 
\author{F. Dominguez}
\affiliation{Institute for Mathematical Physics, TU Braunschweig, 38106 Braunschweig, Germany} 
\author{C. De Beule}
\affiliation{Department of Physics and Materials Science, University of Luxembourg, L-1511 Luxembourg, Luxembourg}
\affiliation{Department of Physics and Astronomy, University of Pennsylvania, Philadelphia PA 19104}
\author{P. Recher}
\affiliation{Institute for Mathematical Physics, TU Braunschweig, 38106 Braunschweig, Germany} 
\affiliation{Laboratory of Emerging Nanometrology, 38106 Braunschweig, Germany}
\date{\today}

\begin{abstract}
Minimally twisted bilayer graphene in the presence of an interlayer bias develops a triangular network of valley chiral modes that propagate along the $AB/BA$ interfaces and scatter at the $AA$ regions. 
The low energy physics of the resulting network can be captured by means of a phenomenological scattering network model, allowing to calculate the energy spectrum and the magnetoconductance in a straightforward way. 
Although there is in general a good agreement between microscopic and phenomenological models, there are some aspects that have not been captured so far with the latter. In particular, the appearance of flatbands in the energy spectrum associated to a localized density of states at the $AA$ regions. 
To bring both approaches closer together, we modify the previous energy independent phenomenological model and add the possibility to scatter to a set of discrete energy levels at the $AA$ regions, yielding a $S$ matrix with energy dependent parameters. Furthermore, we investigate the impact of Coulomb repulsion in these regions on a mean-field level and discuss possible effects of decoherence due to elastic and inelastic cotunneling events.
\end{abstract}
\maketitle

\section{Introduction}

\begin{figure}
	\includegraphics[width=0.45\textwidth]{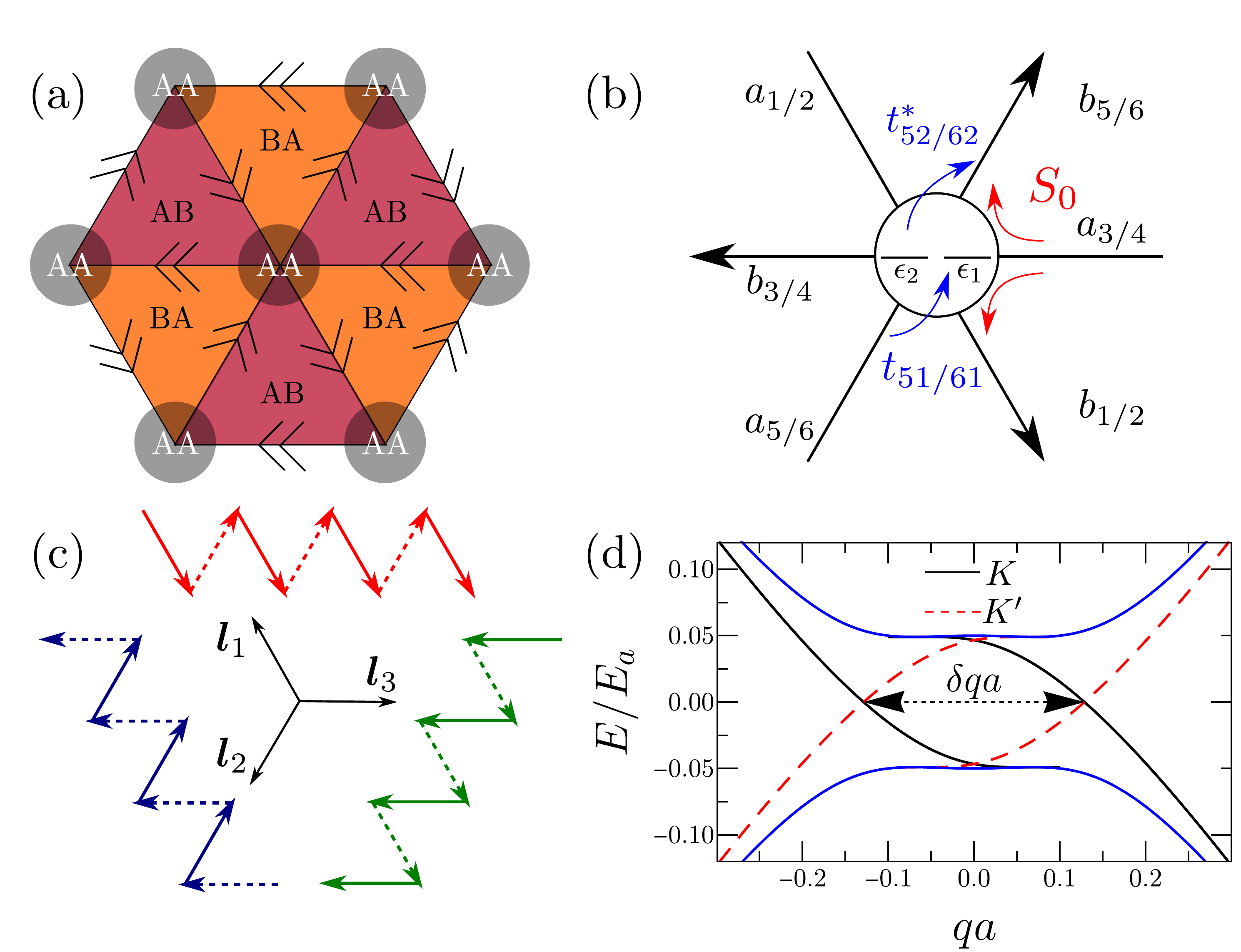}
	\caption{(a) The triangular in mTBG network under an interlayer bias for a given valley and spin. Double arrows represent the two chiral modes propagating along the $AB/BA$ interfaces, which scatter at the $AA$ regions. (b) Sketch of the generalized $S$ matrix for a single $AA$ region. Here, the incoming modes $a_i$ can either deflect to the neighboring outgoing modes $b_i$ via $S_0$ or scatter into the discrete energy levels $\epsilon_j$ from the mode $i$ with the tunneling amplitude $t_{ij}$. (c) Triplet of chiral zigzag modes along the moir\'e lattice vectors ${\bf l}_j$ \cite{DeBeule2021}. (d) Spectrum in units of $E_a=\hbar v_F/a$ of the boundary between two semi-infinite $AB$ and $BA$ regions versus the momentum $q$ along the interface, calculated with the continuum model from Ref. \onlinecite{Zhang2013}. We have used $\Delta/\gamma_\perp=0.1$, with $\Delta$ the interlayer bias and $\gamma_\perp$ the interlayer coupling. We observe two chiral modes per valley (solid black and red dashed lines) copropagating inside the gap of the bulk bands (solid blue).}
	\label{fig.model}
\end{figure}

Twisted bilayer graphene, two stacked graphene layers with a relative twist angle $\theta$ opens the door to control a new degree of freedom in the electronic band structure. Due to the relative twist, the stacking order in the two layers in twisted bilayer graphene varies spatially, resulting in a periodic lattice with a new lattice constant $l=a/2\sin(\theta/2)\approx 14(\theta^\circ)^{-1}\,$nm, the moir\'e lattice constant, with $a$ being the graphene lattice constant \cite{Lopes2007,Morell2010,Li2010,Bistritzer2011}. 
The modulation of the moir\'e lattice constant as a function of $\theta$ can alter significantly the electronic band structure \cite{Lopes2007,Morell2010,Li2010,Bistritzer2011}. 
Indeed, striking results occur at the so-called magic angle $\theta \sim 1^\circ$, where the curvature of the bands is drastically reduced and interacting effects take over, giving rise to a variety of correlated phases \cite{Kim2017,Po2018,Cao2018,Cao2018a,Lu2019,Xie2019,Yankowitz2019,Sharpe2019,Kerelsky2019,Choi2019,Cao2020,Zondiner2020,Wong2020}. 
Beyond the magic angle, new physics arises for small twist angles $\theta \ll 1 ^\circ$,  in a regime called minimally twisted bilayer graphene (mTBG), where a triangular network of valley chiral edge states arise in the presence of an interlayer bias voltage. At such small twist angles, the moir\'e lattice constant becomes very large $l\sim 100\,$nm. Thus, it is energetically more favorable to form a lattice  \cite{Note1} 
consisting of triangular $AB/BA$ Bernal-stacked regions with $AA$ regions at each corner \cite{Prada2013,Nam2017,Walet_2019}, see Fig.~\ref{fig.model}(a). 
Similarly as in Bernal-stacked bilayer graphene\cite{Martin2008,Zhang2013,Vaezi2013}, the presence of an interlayer bias voltage breaks inversion symmetry, yielding a gap opening at the $AB/BA$ stacked regions. 
For a given valley and spin, the difference between the valley Chern number of the $AB/BA$ regions is $\pm 2$. Thus, in the limit of smooth disorder on the scale of $a$ (no intervalley coupling), two chiral modes propagate along the $AB/BA$ interfaces for each valley and spin. 
In this way, mTBG develops a triangular network of valley chiral modes \cite{Prada2013}, which can be visualized experimentally using STM measurements \cite{Ju2015,Yin2016,Huang2018,Sunku2018,Rickhaus2018,Verbakel2021}. 
Remarkably, more refined calculations have shown that far from forming a percolating two-dimensional network, the chiral modes arrange in perfect one-dimensional zigzag (ZZ) channels disposed in three directions \cite{Tsim2020,DeBeule2020}.
This unprecedented scenario of one-dimensional modes propagating in the bulk of the material becomes more apparent in interference experiments. In the presence of a perpendicular magnetic field, effects of Aharonov-Bohm physics arise in the longitudinal and transversal resistance\cite{Xu2019}.

The low energy physics of this system is captured using a phenomenological scattering model that takes into account the symmetries of the system, that is, $C_3$ and $C_2T$, where $C_3$ and $C_2$ are rotations about the axis perpendicular to the plane with respect to the center of an $AA$ region, and $T$ is time reversal symmetry. These symmetries impose the following constraints on the $S$ matrix  
of a single $AA$ region and a given valley and spin\cite{Efimkin2018,DeBeule2020,DeBeule2021}
\begin{align} 
&C_3:\mathcal ~~~~~~~~S_K = \begin{pmatrix} s_f & s_l & s_r \\ s_r & s_f & s_l \\ s_l & s_r & s_f \end{pmatrix},\label{eq:C3}\\
&C_2 T: \qquad s_r =  (s_l)^t, \quad s_f = (s_f)^t,\label{eq:C2}
\end{align}
where the $2\times2$ matrices $s_f$ and $s_{l/r}$ contain the transmission and reflection left/right coefficients of the three pairs of valley chiral modes that arrive at the $AA$ regions. According to Fig.~\ref{fig.model}(b), the $S$ matrix relates the incoming ($a$) and outgoing $(b)$ scattering states via $b=S_K a$ with ${a=(a_1,a_2,a_3,a_4,a_5,a_6)^t}$ and ${b=(b_1,b_2,b_3,b_4,b_5,b_6)^t}$.
Note that the $S$ matrix for the opposite valley $S_{K'}$ is obtained by means of the time-reversal operation ${S_{K'}=(S_K)^t}$.

The simplicity of these phenomenological models allows not only to recover previous results obtained using a microscopic approach\cite{Hou2020, Fleischmann2020, Nguyen_2021}, e.g.~the independent families of chiral ZZ modes, but also to study the topology of the system\cite{DeBeule2021Floquet}, to include electron-electron interactions\cite{Chou2021, park2022network}, effective Bloch oscillations~\cite{Vakhtel2022a} or to calculate two- and four-terminal magnetotransport\cite{DeBeule2020,DeBeule2021Floquet,DeBeule2021}.

There are, however, some features that appear in the spectra of the microscopic calculations that have so far not been captured in these phenomenological models. Here, we will focus on the flatbands placed in the middle of the gap opened by the interlayer bias voltage, which exhibit a localized character at the $AA$ regions. \cite{Prada2013,Ramires2018,Fleischmann2020}
To include these features in the phenomenological approach, we introduce a discrete density of states at the $AA$ regions by means of a collection of $N$ discrete levels which couple to the valley chiral modes propagating along the $AB/BA$ regions [cf. Fig.~\ref{fig.model} (b)]. The resulting $S$ matrix acquires an energy dependence due to the energy difference between the position of the discrete energy levels and the propagating modes. In this context, we investigate the energy spectrum and magnetotransport of the resulting triangular network and analyze the effects of electron-electron interaction on a mean field level at the $AA$ regions, which can lead to Coulomb blockade. We further discuss the impact of decoherence caused by inelastic and elastic cotunneling processes.

The structure of the paper is as follows. In Secs.~\ref{Model} and~\ref{Smatrix}, we introduce the model and the energy-dependent $S$ matrix for a single $AA$ region (node). Then, we calculate the energy bands of the network in Sec.~\ref{Network bands}. In Sec.~\ref{Conductance}, we calculate the two-terminal magnetoconductance of a finite length strip and discuss the impact of cotunneling events that can lead to decoherence.

\section{Model and Hamiltonian for a single $AA$ region}\label{Model}

We model the $AA$ regions taking into account two effects. 
First, we add a set of discrete levels which are coupled to the chiral modes. Second, we also include an energy-independent deflection process between an incoming mode and the neighboring modes, see Fig.~\ref{fig.model}(b). 
This is justified because in the proximity of the $AA$ regions, the chiral modes approach each other, giving rise to a finite overlap between the wave functions. The deflection probability depends on the localization length of the chiral modes relative to the moir\'e length $l$. It is expected that the overlap between neighboring channels decreases if the localization length of the chiral modes perpendicular to the propagation direction is much smaller than the moir\'e wavelength. \cite{Qiao2014, Fleischmann2020}
Furthermore, due to the negligible spin-dependent couplings in graphene systems \cite{KaneMele2005,Min_2006}, such as spin-orbit coupling, we will consider hereafter a spinless Hamiltonian.

We model a single $AA$ region using a set of $N$ discrete energy levels described by means of the Hamiltonian
\begin{align}
	H_{D} &= \sum_{j=1}^N \epsilon_{j} \hat{d}_j^\dagger \hat{d}_j + \frac{U}{2}\sum_{\substack{j,j'\\ j\neq j'}}^N  \hat{n}_j \hat{n}_{j'}\nonumber\\
	&\approx \sum_{j=1}^N \epsilon_{j} \hat{d}_j^\dagger \hat{d}_j + \frac{U}{2} \sum_{\substack{j,j'\\ j\neq j'}}^N  \left(\langle \hat{n}_j\rangle  \hat{n}_{j'}+  \hat{n}_j \langle \hat{n}_{j'} \rangle\right), 
\label{eq:HDL}
\end{align}
with $\epsilon_j$ being the energy of the $j$th level and $U$ is the Coulomb repulsion energy.
Here, $\hat{d}_j^\dagger$ $(\hat{d}_j)$ is the creation (annihilation) operator of the energy level $j$ in the $AA$ region and $\hat{n}_j=\hat{d}_j^\dagger \hat{d}_j$ is the occupation number operator of the $j$th level. 
In the second line of Eq.~\eqref{eq:HDL}, we replace the many-body Coulomb repulsion by a mean-field single-electron term. This approximation is valid as long as the fluctuations in the occupation number are small. 
\cite{Note2}
In this mean-field description we cannot capture all effects that can possibly occur, such as cotunneling processes that change the occupation of the discrete levels. As we will argue at the end of Sec.~\ref{Conductance}, these processes can lead to decoherence of interference effects. We further assume that the discrete levels are in equilibrium with the external leads of the network, which set the Fermi energy $E_F$.
Details of the self-consistent calculation of $\langle \hat{n}_i\rangle$ are given in App.~\ref{App.Selfcon}.

We include possible degeneracies present in the $AA$ regions and set a two-fold degenerate level $\epsilon_{j}=\epsilon_{j+1}$. We can motivate these degeneracies by spin or orbital degrees of freedom, which can arise for example in bilayer graphene quantum dots \cite{Recher2009, Recher2010}. Furthermore, we consider that the energy difference between the nearest non-degenerate levels is the largest energy scale in the problem. In this way, we reduce the number of relevant discrete levels per node to two, i.e.~$N=2$ and $\epsilon:=\epsilon_{1}=\epsilon_{2}$. Note, however, that the extension to an arbitrary number of states is straightforward.

The coupling between the chiral modes and the discrete energy levels at position $x_{AA}$ is given by the tunneling Hamiltonian 
\begin{align}
H_T = \sum_{i=1}^6 \sum_{j=1}^N t_{ij}\hat{d}^\dagger_j \hat{\psi}_i(x_{AA})+h.c.\label{VI}, 
\end{align}
where $t_{ij}$ describes the tunnel amplitudes between the $i$th chiral mode and the $j$th discrete energy level. $C_3$ symmetry imposes $t_{1,j}=t_{3,j}=t_{5,j}$ and $t_{2,j}=t_{4,j}=t_{6,j}$. If in addition $C_2T$ symmetry is fulfilled, then $t_{1,j}$ and $t_{2,j}$ are real.
The field operators $\hat{\psi}(x)$ account for the chiral modes, which propagate due to the kinetic energy term 
\begin{equation}
H_{0} = \sum_{i=1}^6 \int_{-\infty}^\infty dx_i \hat{\psi}_i^\dagger(x_i)(-i\hbar v_i) \partial_{x_i}\hat{\psi}_i(x_i), \label{eq.H0}
\end{equation} where $x_i$ is the coordinate of the $i$th chiral mode with velocity ${|v_i|=v_F}$ and $v_F$ is the fermi velocity of graphene.

\section{Generalized $S$ matrix}\label{Smatrix}
 
To include the deflection processes together with the coupling to the discrete levels into a single $S$ matrix, we make use of the generalized Weidenm\"uller formula calculated by means of the equation of motion method \cite{Hackenbroich2001,Aleiner2002,Tripathi2016}
\begin{align}
S(E) = &S_0 -i \pi \nu \left(1+i \pi \nu R\right)^{-1}\mathcal{W}^\dagger [E-h_D \nonumber\\
&+i \pi \nu \mathcal{W} \left(1+i\pi \nu  R\right)^{-1}\mathcal{W}^\dagger]^{-1}\mathcal{W}(1+S_0)\label{eq.MWhere},
\end{align}
where $\nu=1/\pi \hbar v_F$ and $R$ is a decomposition of the energy independent $S$ matrix $S_0$ given by
\begin{equation}
S_0 = \left(1+i\pi \nu R\right)^{-1}\left(1-i\pi \nu R\right). \label{eq.S0}
\end{equation}
The explicit form of $R$ is given in App.~\ref{App.detailsTmatrix}.
Moreover, the matrix $\mathcal{W}$ accounts for the coupling between the chiral modes and the discrete energy levels and $h_D$ is a matrix representation of the Hamiltonian in Eq.~\eqref{eq:HDL}. Without the deflection processes ($R=0$), Eq.~\eqref{eq.MWhere} reduces to the well known Mahaux-Weidenm\"uller formula\cite{Mwformula68}
\begin{align}
S_D = \mathds{1} -2i \pi \nu \mathcal{W}^\dagger\left[E-h_D+ i \pi \nu \mathcal{W} \mathcal{W}^\dagger\right]^{-1}\mathcal{W}. \label{eq:SD}
\end{align} 
We now introduce $S_0$ and $S_D$ separately.

\subsection{Deflection processes $S_0$} \label{PhenoSmatrix}
We model the deflection processes taking place between the incoming mode and the neighboring outgoing modes using the phenomenological $S$ matrix from Refs.~\onlinecite{DeBeule2020,DeBeule2021}, see Fig.~\ref{fig.model}(b). The parameterization of $S_0$ with forward scattering under $C_3$ and $C_{2}T$ symmetries is given by 
\begin{align}
s_f & = \begin{pmatrix} e^{i(\phi+\chi)}\sqrt{P_{f1}} & -\sqrt{P_{f2}} \\ -\sqrt{P_{f2}} & -e^{-i(\phi+\chi)}\sqrt{P_{f1}} \end{pmatrix}, \label{eq:sf} \\
s_r & = \begin{pmatrix} e^{i\phi}\sqrt{P_{d1}} & \sqrt{P_{d2}} \\ -\sqrt{P_{d2}} & -e^{-i\phi}\sqrt{P_{d1}} \end{pmatrix}, \label{eq:sr}
\end{align}
with $s_l=(s_r)^t$ and where $P_{f1}$ ($P_{f2}$) is the probability for intrachannel (interchannel) forward scattering, and $P_{d1}$ ($P_{d2}$) is the probability for intrachannel (interchannel)  deflections. The phase $\phi$ is an independent real parameter and accounts for a phase difference picked up in the scattering process. This parameter tunes the network from decoupled ZZ chiral modes propagating in three different directions ($\phi=0)$, see Fig.~\ref{fig.model} (b), to flatbands performing closed orbits around the $AB$ and $BA$ regions ($\phi=\pi/2$). The $S$ matrix is unitary for ${2(P_{d1}+P_{d2})+P_{f1}+P_{f2}=1}$ and ${\cos \chi = \left( P_{d2} - P_{d1} \right)/2\sqrt{P_{f1} P_{d1}}}$, where we take $\chi \geq 0$. Moreover, $\chi$ has to be real which implies ${2\sqrt{P_{f1}P_{d1}} \geq \left| P_{d2} - P_{d1} \right|}$. Since we only consider deflections here, we set $s_f=0$ and $P_{d1}=P_{d2}$, such that $\chi=\pi/2$.

\subsection{Discrete levels $S_D$}

The matrix $h_D$ entering in Eqs.~\eqref{eq.MWhere} and~\eqref{eq:SD} is given by
\begin{align}
h_D = \begin{pmatrix}\epsilon_{1} + U \langle \hat{n}_2\rangle & 0 \\ 0 & \epsilon_{2} + U \langle \hat{n}_1 \rangle 
       \end{pmatrix},
\end{align}
which is obtained from Eq.~\eqref{eq:HDL}, with $N=2$, that is, $H_D = (\hat{d}_1^\dagger,\hat{d}_2^\dagger)h_D(\hat{d}_1,\hat{d}_2)^t$.

From Eq.~\eqref{VI}, we write 
\begin{equation}
H_T=(\hat{d}_1^\dagger,\hat{d}_2^\dagger) \mathcal{W} (\hat{\psi}_{1},\hat{\psi}_{2},\hat{\psi}_{3},\hat{\psi}_{4},\hat{\psi}_{5},\hat{\psi}_{6})^t+h.c.
\end{equation}
with
\begin{equation}
	\mathcal{W}= \begin{pmatrix}
	t_{1,1} & t_{2,1} & t_{1,1} & t_{2,1}& t_{1,1} & t_{2,1}\\
	t_{1,2} & t_{2,2} & t_{1,2} & t_{2,2}& t_{1,2} & t_{2,2}\\
	\end{pmatrix}, 
\end{equation}
where $t_{i,j}\in \mathds{R}$ for $i,j=1,2$.
For simplicity, we impose that any given valley chiral mode couples with the same amplitude to every discrete state, yielding a single tunnel amplitude $t_{D}:=t_{i,1}=t_{i,2}$. As mentioned already in Sec.~\ref{Model}, we will stick to the case of $\epsilon:=\epsilon_1=\epsilon_2$. Therefore, the energy difference between the two levels is caused by the Coulomb repulsion energy $U$.

The expression for $S_D$ with the previous mentioned simplifications is given in App.~\ref{App.details1p1}. We also show that the scattering matrix $S_D$ for the case with a single tunnel amplitude $t_\text{D}$ can be written as three free chiral modes plus the one channel Efimkin-MacDonald model \cite{Efimkin2018}.

\begin{figure*}
	\includegraphics[width=0.9\textwidth]{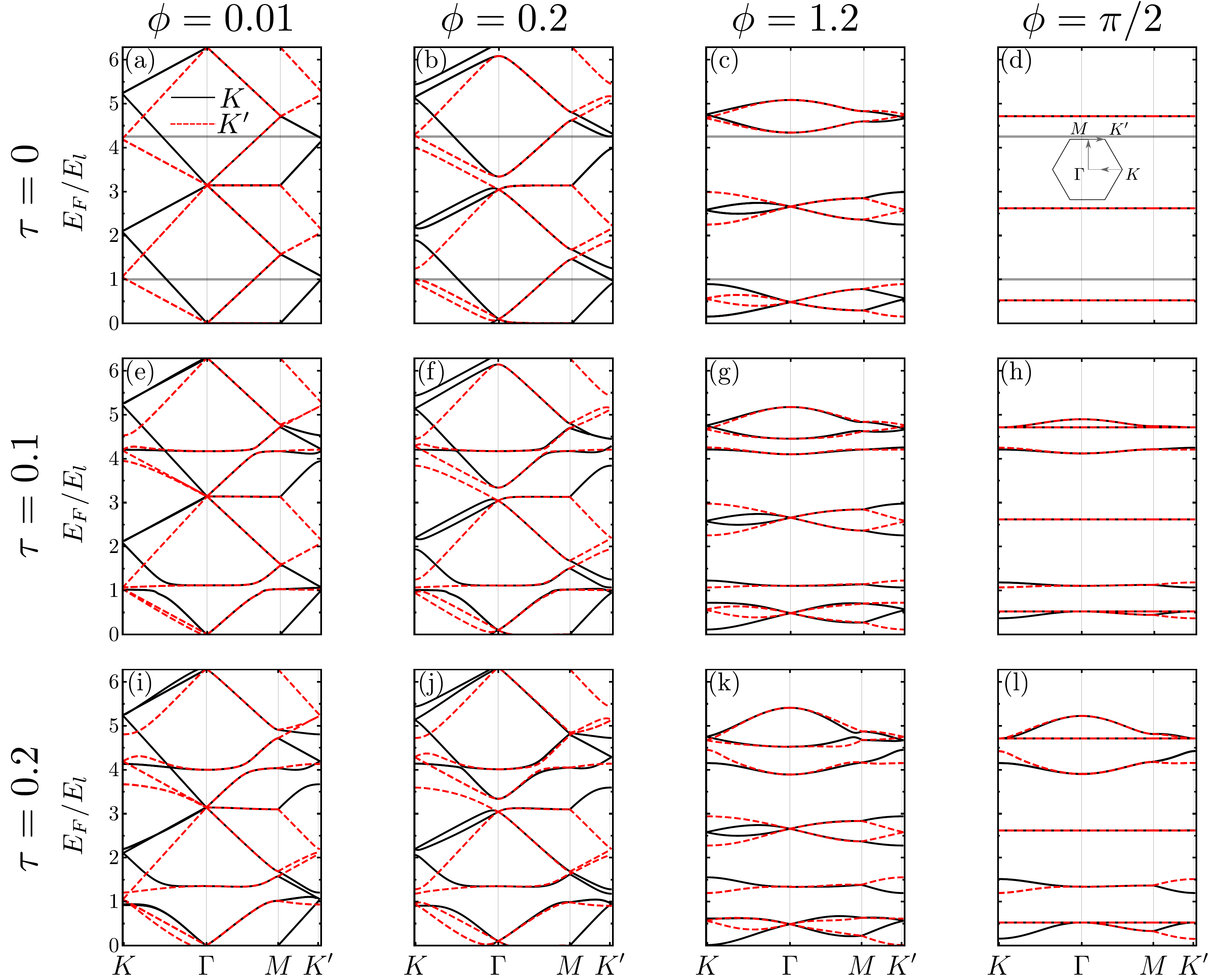}
	\caption{Network bands along high symmetry lines, see inset of panel (d), for different values of $\phi$ and $\tau=t_{D}\sqrt{l}/\hbar v_F$ in units of $E_l=\hbar v_F/l$. Black solid lines belong to the $K$ valley and red dashed lines to the $K'$ valley. Horizontal grey lines indicate the position of the energy levels $\epsilon=E_l$ and $\epsilon+U=4.25E_l$, $\tau=0$ for [(a)-(d)], $\tau=0.1$ for [(e)-(h)] and $\tau=0.2$ for [(i)-(l)].} \label{bandspicture}
\end{figure*}

\newpage 

\section{Network energy bands} \label{Network bands}

We now construct a triangular network of chiral modes making use of the $S$ matrix introduced above and Bloch's theorem, see Refs.~\onlinecite{Efimkin2018, DeBeule2020, DeBeule2021}. To this aim, we place the $AA$ regions (nodes) at positions $m \bm l_1 + n \bm l_2$, with $m,n \in \mathbb{Z} $  and $\bm l_{1,2} = l (-1/2,\pm \sqrt{3}/2)$ are moir\'e lattice vectors, see Fig.~\ref{fig.model}(c). We relabel the incoming scattering amplitudes as

\begin{equation}
a_{mn} = \left( a_{1mn}, a_{2mn}, a_{3mn}, a_{4mn}, a_{5mn}, a_{6mn} \right)^t,
\end{equation}
and $b_{mn}$ for the outgoing scattering states. In this notation, the $S$ matrix given in Eq.~\eqref{eq.MWhere} becomes, 
\begin{equation}
    b_{mn} = S(E) a_{mn}. \label{eq.sbands}
\end{equation}

Furthermore, the incoming states at the node $(m,n)$ are the outgoing states of neighboring nodes, more specifically
\begin{align}
a_{mn} & = e^{i\frac{E }{E_l}} \left( b_{1m+1n}, b_{2m+1n}, b_{3m-1n-1},\right. \nonumber \\
& \hspace{1.8cm} \left. b_{4m-1n-1}, b_{5mn+1}, b_{6mn+1} \right),\label{eq.dynphase}
\end{align}
where we include the effects of a dynamical phase picked up along the propagation between nodes. We use $E_l=\hbar v_F/l$ as the energy scale of the system.

Making use of the translational invariance, Bloch's theorem relates \cite{Efimkin2018, Pal2019}
\begin{equation} \label{eq:bloch}
\begin{pmatrix} b_{1m+1n} \\b_{2m+1n} \\ b_{3m-1n-1}\\ b_{4m-1n-1} \\ b_{5mn+1} \\ b_{6m,n+1}\end{pmatrix}_{\bm k} = \left[ \mathcal M(\bm k) \otimes \mathds 1_2 \right] \begin{pmatrix} b_{1mn} \\ b_{2mn} \\ b_{3mn} \\ b_{4mn} \\ b_{5mn} \\ b_{6mn} \end{pmatrix}_{\bm k},
\end{equation}
with $\mathcal M(\bm k) = \textrm{diag} \left( e^{ik_1}, e^{ik_3}, e^{ik_2} \right)$ where $k_j = \bm k \cdot \bm l_j$ ($j=1,2,3$) and $\bm l_3 = -(\bm l_1+\bm l_2)$. Finally, we substitute Eq.~\eqref{eq.sbands} and Eq.~\eqref{eq.dynphase} into Eq.~\eqref{eq:bloch}, and find
\begin{align} \label{bands}
\left[ \mathcal M(\bm k) \otimes \mathds 1_2 \right] S(E) \, a_{\bm k} = e^{-i\frac{E }{E_l}} a_{\bm k},
\end{align}
whose solutions gives the energy spectrum of the system. 
In the limit of $t_{D}\rightarrow 0$, the $S$ matrix is independent of $E$ and the spectrum is obtained taking $i \log(\xi_j)$, with $\xi_j$ being the $j$th eigenvalue of $\left[ \mathcal M(\bm k) \otimes \mathds 1_2 \right] S$. The resulting spectrum is periodic in energy with period $2 \pi \hbar v_F/l$, which can be inferred from Eq.~\eqref{bands}. 
In turn, for finite coupling $t_D\neq 0$, the $S$ matrix depends on $E$, and thus, the spectrum is obtained by solving
\begin{equation}
\text{det}\left([\mathcal{M}(\bm k)\otimes \mathds{1}_2]S(E)-e^{-i \frac{E }{E_l}}\mathds{1}_6 \right) = 0,
\end{equation}
which is highly nonlinear and contains an infinite set of solutions for a given $(k_x,k_y)$. In this situation, the energy spectrum becomes periodic only far from the resonances, where the energy dependence of $S$  is surpressed.
Note that the energy spectra for the opposite valley is obtained by using $S_{K'}$ and reversing the sign of the momenta introduced in $\mathcal{M}(\bm k)\rightarrow \mathcal{M}(-\bm k)$.

\subsection{Relation between $\delta q$ and $\phi$} \label{Sec.dq}

Before analyzing the numerical results of the energy spectrum, it is important to pay attention to the momentum difference $\delta q$ exhibited by the chiral modes copropagating along a single $AB/BA$ interface \cite{Zhang2013}, see Fig.~\ref{fig.model} (d). Similarly as with the dynamical phase accumulated between two consecutive $AA$ regions in Eq.~\eqref{eq.dynphase}, the effects of the momentum difference are included by multiplying the scattering amplitudes by $Z=\exp[i (\delta q l/2)  \mathds{1}_3\otimes\sigma_z]$,  where $\sigma_z$ is the Pauli matrix in the space of copropagating chiral modes. When the chiral modes scatter into an $AA$ region, they become mixed, and thus, the total phase of each outgoing mode can be compensated or accumulated.
The addition of this phase difference modifies the network spectrum in the same way as $\phi$ does\cite{DeBeule2020,DeBeule2021}. To see this, we add the phase $(\delta q l/2)  \mathbb{1}_{3}\otimes \sigma_z$ to the dynamical phase in Eq.~\eqref{eq.dynphase} and obtain in this case the following equation for the energy spectrum
\begin{equation}
\text{det}\left([\mathcal{M}(\bm k)\otimes \mathds{1}_2]S'(E)-e^{-i \frac{E}{E_l} }\mathds{1}_6 \right) = 0,
\end{equation}
with $S'(E)=Z S(E)$. Here, we used that the matrix $\mathcal{M}(\bm k)\otimes \mathds{1}_2$ commutes with $Z$. 

The $S$ matrix found in Refs.~\onlinecite{DeBeule2020,DeBeule2021}, as well as, the one derived in Eq.~\eqref{eq.MWhere} without the discrete levels ($t_D=0$), satisfies $Z S_0(\phi_0)=M S_0(\phi_0+\phi) M^\dagger$, with $M = \sqrt{Z} = \exp[i (\delta q l/4)  \mathds{1}_3\otimes\sigma_z]$ for $\phi=\delta ql/2$, which demonstrates the equivalence $\phi=\delta q l/2$. This means that there are two microscopic origins of the phase difference $\phi$: It can be accumulated along the links due to the momentum difference of the chiral modes or directly by scattering with the $AA$ region.

Assuming that the phase $\phi$ is entirely picked up along the links, we estimate $\phi$ in the experiment reported in Ref.~\onlinecite{Xu2019}. There, the electric field opened a gap around $\sim 50\,$meV and the interlayer coupling $\approx 0.39\,$eV yield a momentum difference of $\delta q\approx 0.17 \,$nm$^{-1}$. Now, the twist angle gives rise to $l\approx 140\,$nm, hence, $\phi'=\delta q l/2\approx 3.89 \pi$ and $\phi=4\pi-\phi'\approx0.34$. This value of $\phi$ sets the system into the chiral ZZ regime, where the three families of chiral ZZ modes are weakly coupled and it is in accordance with the analysis performed in Ref.~\onlinecite{DeBeule2020}.

\subsection{Numerical results}

We are now ready to calculate the energy spectrum in the presence of discrete energy levels at the $AA$ regions. We consider different values of $t_\text{D}$ and $\phi$. We introduce the dimensionless parameter 
\begin{equation}
    \tau = t_D \sqrt{l}/\hbar v_F,
\end{equation}
to quantify the coupling $t_D$ with respect to the kinetic energy of the valley chiral modes. For the scattering parameters of the deflecting scattering matrix $S_0$, we use $P_{d1}=P_{d2}=1/4$.
Furthermore, the parameter $\phi$ enters as the phase factor picked up along the links of the network, and therefore, it affects both $S$ matrices $S_0$ and $S_D$, see Sec. \ref{Sec.dq}. We assume that the occupation of the discrete levels is determined by equilibrium with the external leads, which set the Fermi energy $E_F$. For this reason, we give our results as a function of $E_F$.

For $\tau=0$, we reproduce the network bands of the phenomenological scattering matrix\cite{DeBeule2020,DeBeule2021}, see Fig.~\ref{bandspicture} panels (a)-(d). For $\phi \approx 0$ [panel (a)] the spectrum disperses linearly resulting from the presence of ZZ chiral modes propagating in three different directions in the bulk of the material \cite{Tsim2020,DeBeule2020,DeBeule2021}. Increasing slightly $\phi=0.2$ [panel (b)] gives rise to a finite coupling process between the chiral ZZ modes. Consequently, indirect gaps open in the spectrum. For larger values of $\phi=1.2>\pi/6$ [panels (c) and (d)], the ZZ bands hybridize further and loose their chiral character, yielding the opening of gaps in the spectrum.

Next, we set a finite coupling $\tau\neq 0$. We differentiate between two energy sectors, far from resonance, i.e.~${|E_F-\epsilon| \gg |\tau|^2 E_l}$ and ${|E_F-\epsilon-U| \gg |\tau|^2 E_l}$, and close to resonance. In the former limit, the system becomes effectively described by the deflection matrix $S_0$. Consequently, electrons do not tunnel through the discrete energy states. Therefore, the network bands for finite coupling are far from the discrete energy levels remain the same as without coupling to the discrete levels. Close to resonance ${|E_F-\epsilon| \lesssim |\tau|^2 E_l}$ or ${|E_F-\epsilon-U| \lesssim |\tau|^2 E_l}$, the $S$ matrix becomes a combination of both $S$ matrices, $S_0$ and $S_D$. Analyzing the network energy spectrum at resonance, that is by varying simultaneously $E_F = \epsilon$, we observe high and low conductance regimes, see App.~\ref{App.res}. Therefore, depending on the values of $\epsilon$ and $\epsilon+U$, we will observe a different response close to resonance. To explore both scenarios, we analyze the case in which $\epsilon$ and $\epsilon+U$ coincide with a low and a high conductance limit. Hence, we will use for the calculations $\epsilon=\hbar v_F/l$ and $U=3.25\hbar v_F/l$.

In the network bands, the discrete energy levels appear as flatbands [panels (e)-(l)]. For higher values of $\tau$ the energy window, where the influence of the discrete energy levels is dominant, becomes bigger and therefore the bandwidth of the nominal flatbands grows with larger $\tau$ [panels (i)-(l)].

\section{Magnetoconductance} \label{Conductance}

\begin{figure}
	\includegraphics[width=0.5\textwidth]{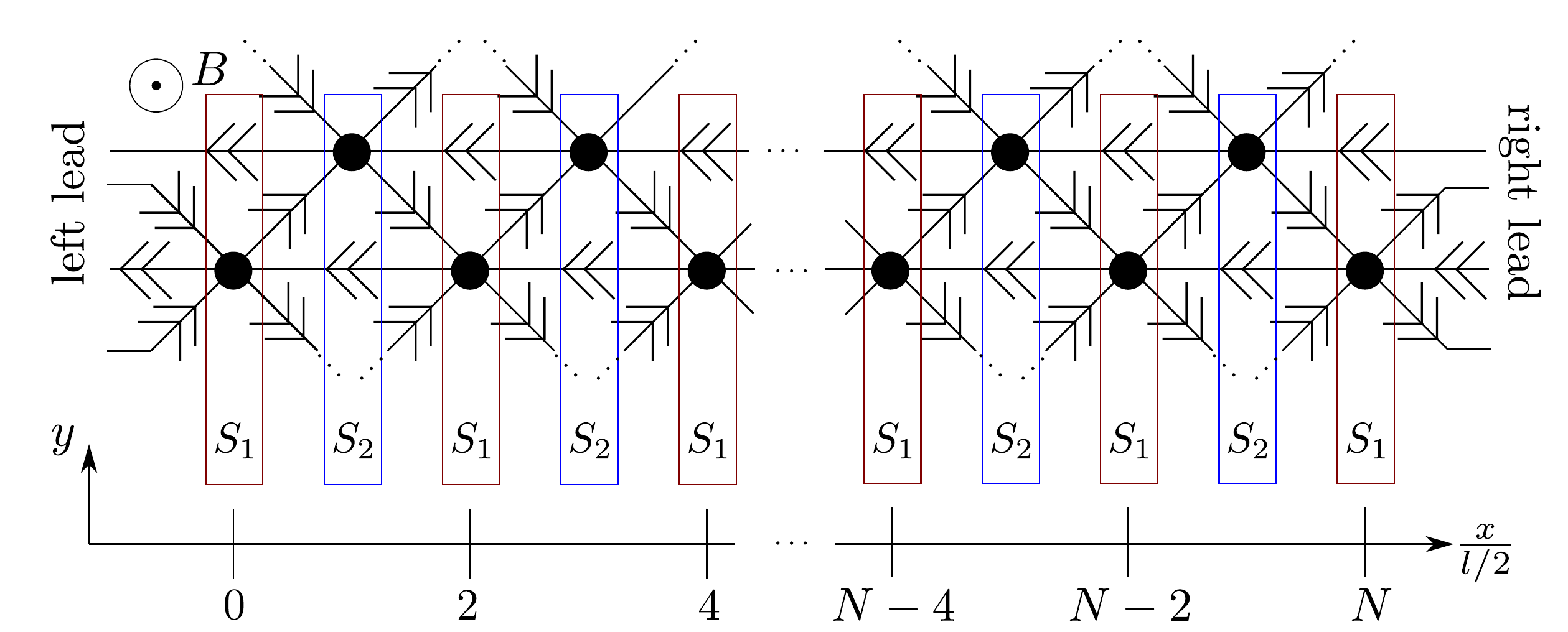}
	\caption{Network strip with length $L=Nl$ and width $W\gg L$. Black circles depict the scattering $AA$ regions and $S_1$, $S_2$ represent the inequivalent scattering sections used to calculate the conductance.}
	\label{fig.Networkmodel}
\end{figure}

We consider a wider-than-long strip in contact to left and  right leads and calculate the conductance as a function of the Fermi energy $E_F$ in the presence of a magnetic field $B$ perpendicular to the plane of the system. 
To this aim, we make use of Bloch's theorem in the transverse direction describing an infinitely wide strip and set a finite length in the longitudinal direction, see Fig.~\ref{fig.Networkmodel}. Here, particles can acquire three different phases when propagating from one node to the neighboring one, that is, the phase $\phi=\delta q l/2$ due to the momentum difference, a dynamical phase $\phi_{\text{dyn}}= E/E_l$ with $E_l=\hbar v_F/l$ and the Peierls phase 
\begin{equation}
	\Phi_P\left(\frac{x}{l/2}\right)=\frac{\pi \Phi}{\Phi_0} \left(\frac{x}{l/2}+\frac12\right),
\end{equation}
which accounts for the effects of the applied magnetic field $B$ and gives rise to the magnetic flux $\Phi = B \mathcal{A}$ through a moir\'e unit cell with area $\mathcal{A}= (\sqrt{3}/2)l^2$, where $\Phi_0 = e/h$ is the magnetic flux quantum. The phase $\pm\Phi_P$ is accumulated starting from a node with horizontal position $x$ in downward/upward direction. Here, we have used the Landau gauge $\mathbf{A}=Bx \mathbf{e}_y$.

\begin{figure}
\includegraphics[width=0.5\textwidth]{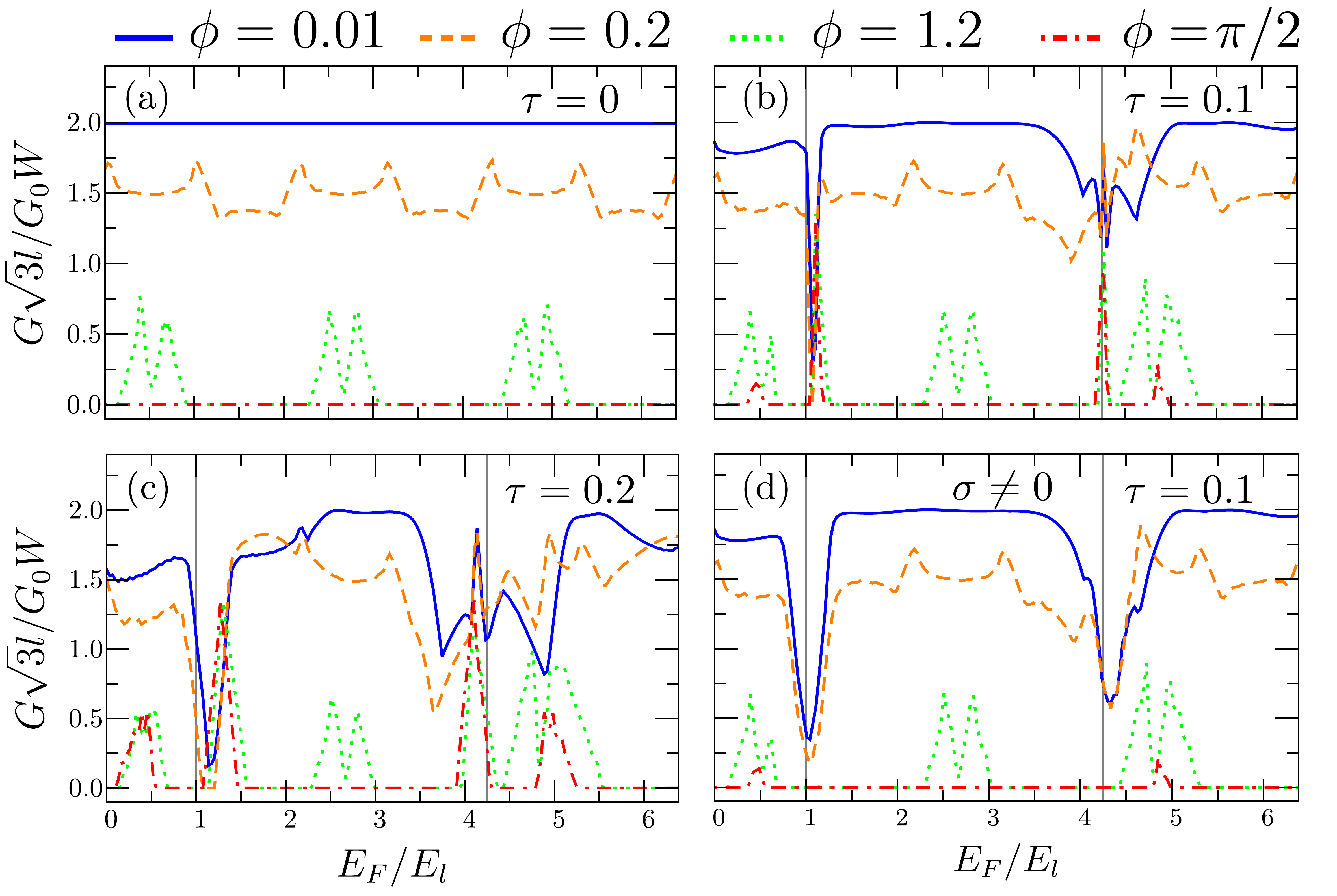}
	\caption{Conductance as a function of $E_F$ for a network model with length $L=20l$ and $W\gg L$ for different $\phi$ and couplings to the $AA$ region $\tau=t_{D}\sqrt{l}/\hbar v_F$ over $E_F$ in units of $E_l=\hbar v_F/l$ and temperature $T=0$. The other parameters are $\epsilon=E_l$, $U=3.25E_l$, $G_0=4e^2/h$. In (d) instead of fixed energy levels $\epsilon_{1}$, $\epsilon_{2}$ we have used a normal distribution with mean value $\epsilon_{\mu} =E_l$ and $U =3.25E_l$ and a standard deviation of $\sigma=\hbar v_F/10l$ from which the values of the energy levels were randomly picked for each network strip.} \label{Cond}
\end{figure} 

We calculate the two terminal linear conductance by means of the Landauer-Büttiker formalism \cite{Buettiker1986,Buettiker88} 
\begin{equation}
G = \frac{4e^2}{h} \frac{W}{\sqrt{3}l}\int dE  \mathcal{T}(E) \left(-\frac{\partial f_0(E)}{\partial E}\right), \label{eq.Cond}
\end{equation}
where $W$ is the width of the strip, $l$ is the moir\'e lattice constant and the factor 4 accounts for the spin and valley contributions. Here, $f_0$ is the Fermi-Dirac distribution given by
\begin{equation}
f_0(E) = \frac{1}{\exp\left[(E-E_F)/k_B T\right]+1},
\end{equation}
and $E_F$ is the Fermi energy. Furthermore, $\mathcal{T}(E)\in[0,4]$ is the transmission function per unit cell for a given valley and spin, which is calculated by the recursive combination of $S$ matrices corresponding to sections along the strip, see Fig.~\ref{fig.Networkmodel}. Further details of the transport calculations are presented in App.~\ref{App.Transport}.

\begin{figure}[tb]
	\includegraphics[width=0.5\textwidth]{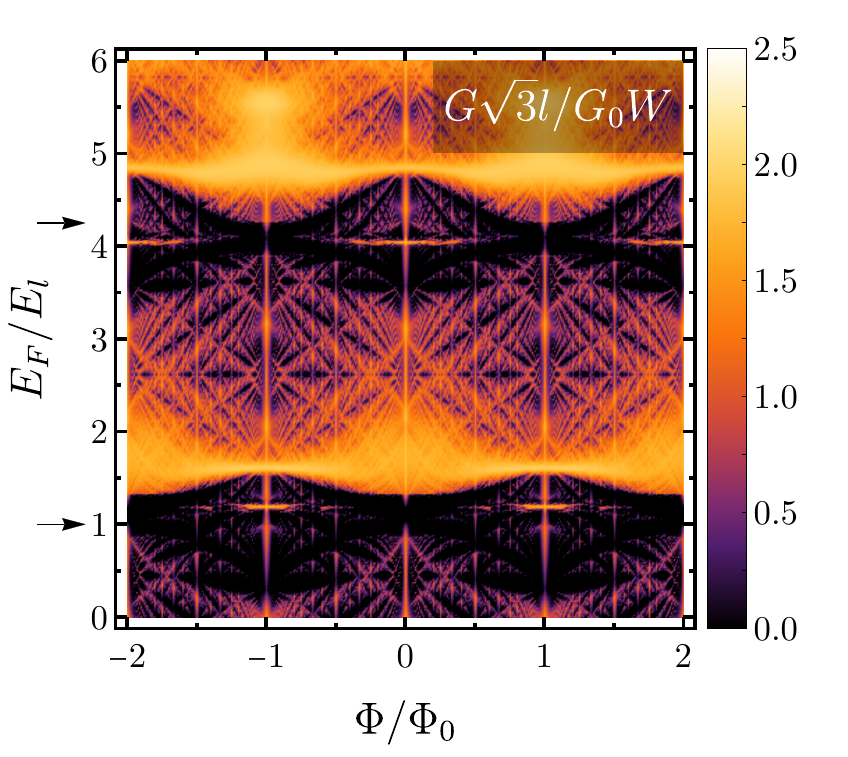}
	\caption{Conductance at zero temperature as a function of $E_F$ and the magnetic flux $\Phi$ for a network with length $L=20l$ and $W\gg L$. We have used $\tau= 0.2$, $\epsilon=E_l$, $U=3.25E_l$, $\phi=0.2$, $G_0=4e^2/h$ and $E_l=\hbar v_F/l$. We highlight the lower energy level at $\epsilon$ and the upper energy level at $\epsilon+U$ with black arrows at the vertical axis.} \label{CondDensity}
\end{figure}

\begin{figure*}[tb]
	\includegraphics[width=\textwidth]{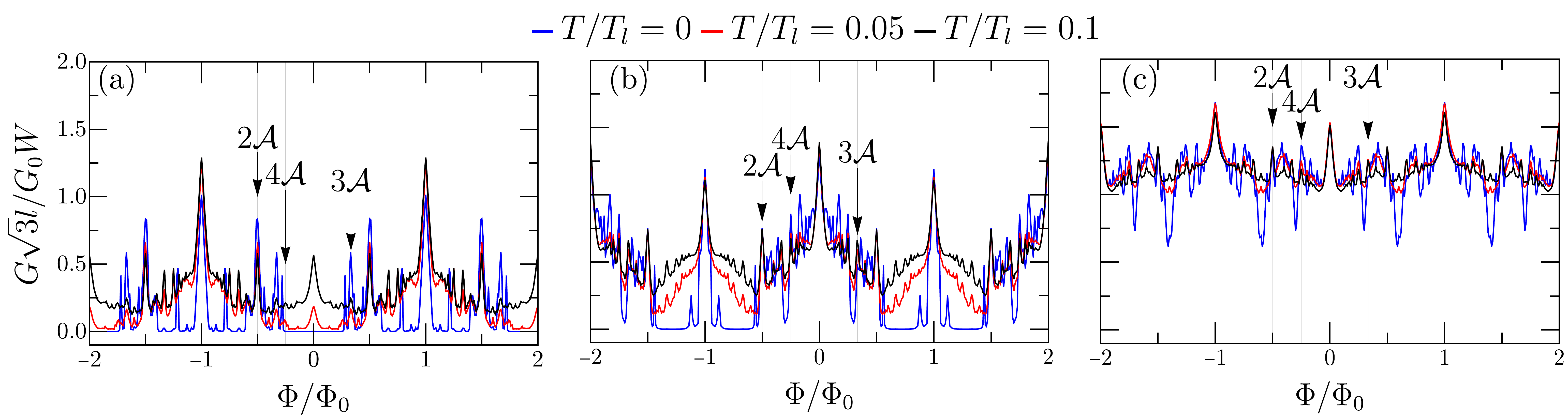}
	\caption{Conductance as a function of the magnetic flux $\Phi$ for a network with length $L=10l$ and different temperatures. In (a) we show the magnetoconductance close to the lower energy level ($E_F =1.1E_l$), in (b) close the higher energy level ($E_F=4.24 E_l$) and (c) far from the energy levels ($E_F =20\hbar v_F/l$) for different temperatures. The temperature scale is $T_l = \hbar v_F/(k_B l)$ and the energy scale $E_l= \hbar v_F/l$. We have used $\tau =0.2 $, $\epsilon=E_l$, $U=3.25 E_l$, $G_0=4e^2/h$ and $\phi=0.2$. Encircling half of the moir\'e unit cell $\mathcal{A}/2$ $n$ times leads to an Aharonov-Bohm resonance. Certain values of $n$ are highlighted.} \label{CondTemp}
\end{figure*}

{\bf Conductance at $\mathbf{B=0}$}---
The conductance per unit cell at zero temperature for a network strip of length $L$ and width $W\gg L$ is 
\begin{equation}
\frac{G}{G_0} = \frac{W}{\sqrt{3}l} \mathcal{T}(E_F), \label{eq.condunit}
\end{equation}
with $G_0=4e^2/h$.
In Fig.~\ref{Cond}, we show the conductance as a function of the Fermi energy $E_F$ for three different tunnel amplitudes $\tau$ and four values of $\phi$. For $\tau=0$, we recover the results of the phenomenological model with $P_f=0$ shown in Refs.~\onlinecite{DeBeule2020,DeBeule2021}, see panel~(a).
As we have discussed above, for $\phi\approx0$, the system is metallic and composed of three families of chiral ZZ modes, which give an energy independent conductance $G\sqrt{3}l/G_0W=2$. 
Furthermore, for $\phi= 0.2$, the ZZ chiral modes start hybridizing, leading to the appearance of indirect gaps, around which the conductance develops a peaked structure. Since the linear conductance follows a similar structure as the density of states, we expect to observe maxima in the conductance around the indirect gaps due to the presence of van-Hove singularities characteristic of one-dimensional systems. For larger values of $\phi=1.2 \lesssim \pi/2$, the system develops gaps, yielding zero conductance at those positions.
Finally, for $\phi\approx \pi/2$, the conductance is zero since the electrons tend to form closed orbits around the $AB$ and $BA$ regions \cite{DeBeule2020,DeBeule2021}. 

For a finite coupling to the discrete levels, $\tau \neq0$, the conductance exhibits appreciable changes around the resonances. As we have seen above, the band structure is modified around the position of the discrete levels. For small $\phi\approx 0$, the discrete levels exhibit a flatband dispersion and consequently, close to the resonances the conductance exhibits dips. The width of these dips are proportional to $|\tau|^2\hbar v_F/l$ (cf. Sec.~\ref{Smatrix} C) for small coupling. As we have mentioned above, we expect to observe a different behavior at $E_F=\epsilon$ and $E_F=\epsilon+U$, since they are placed in different regimes of the model at resonance, see App.~\ref{App.res}. Indeed, we observe a larger dip width for $E_F=\epsilon+U$, developing a peak exactly at resonance, see Fig.~\ref{Cond}. For larger values of $\pi/6\leq \phi\leq\pi/2 $, the coupling to the discrete levels adds some curvature to the otherwise flat bands, increasing the conductance. Note, however, the conductance peaks within the network gap [$E_F\sim \epsilon$ and $E_F\sim \epsilon+U$ in Fig.~\ref{Cond} (d)] come from a spurious effect from considering an energy-independent coupling. A more accurate approach would be to calculate the coupling self-consistently, considering the energy-dependent density of states of the network. Then, the coupling would be zero if the energy levels lie inside a gap of the network bands, and therefore, there would be no conductance peaks, see more details in the appendix App.~\ref{App.Selfcon}. 

The impact of the presence of the discrete levels can be slightly different if instead of considering a fixed value of $\epsilon$, we take a random distribution. This distribution is defined by a mean value $\epsilon_\mu$ and a standard deviation $\sigma$. We average over 20 random samples.  Note that we are still looking at a network strip, so the periodicity in the transversal direction is not broken. Doing so, we observe in Fig.~\ref{Cond}(d) that the width of the dips observed for $\phi\approx 0$ increases with respect to panel~(b) and becomes determined not only by $\tau$ but also by the variance of the energy distribution. Also the small peak structure at the higher energy level is no longer visible. Furthermore, the spurious conductance peaks developed at $\phi\approx \pi/2$ become suppressed.  

{\bf Conductance at $\mathbf{B\neq0}$}--- We now include the effects of the magnetic flux $\Phi$ and study the conductance as a function of both $E_F$ and $\Phi/\Phi_0$ for $T=0\,$K, $\phi=0.2$ and $\tau=0.2$, see Fig.~\ref{CondDensity}. Here, we observe two low conductance regions for energies close to $E_F=\epsilon= E_l$ and $E_F=\epsilon+U= 4.25E_l$. Since we have set $\phi=0.2$ (weakly coupled chiral ZZ modes), the system is highly conducting away from the resonances. Then, close to the resonances, the conductance is reduced. Furthermore, the density plot develops a fractal structure, also known as the Hofstadter pattern \cite{Hofstadter1976}. Here, the unusually big moir\'e unit cell allows to observe this structure for experimentally realizable magnetic fields \cite{DeBeule2021}.

The periodicity of the conductance as a function of $\Phi/\Phi_0$ contains information about the underlying network physics. In general, the magnetoconductance shows Aharonov-Bohm (A-B) resonances at multiples of ${\Phi=2\Phi_0/n}$, with $n$ being integer, which originate from trajectories that encircle an area $n\mathcal{A}/2$ in the presence of a magnetic flux, see Fig.~\ref{CondTemp}. When, for example, only parallel chiral ZZ modes are coupled, the periodicity of the conductance becomes $\Phi/\Phi_0= 1$, because the minimal area that particles propagating along the network can encircle is $\mathcal{A}$.\cite{DeBeule2020} This also occurs in the special case of $P_f=0$ and $P_{d1}=P_{d2}$, that is, far from resonance. In this case, $S_0$ allows to couple non-parallel ZZ chiral modes, but again, the minimal area that particles can encircle is $\mathcal{A}$.
In turn, when all possible couplings between ZZ chiral modes are allowed, particles can encircle half the unit cell, yielding a larger periodicity $\Phi/\Phi_0=2$.
To see this more explicitly, we calculate separately the conductance as a function of $\Phi/\Phi_0$ at three different values of $E_F$: close to the resonances $E_F=1.1\epsilon$ [Fig.~\ref{CondTemp} (a)] and $E_F\approx 0.997 (\epsilon+U)$ [Fig.~\ref{CondTemp} (b)], where the periodicity is recovered after $\Phi=2\Phi_0$. While far from resonance $E_F=20 \epsilon$ [Fig.~\ref{CondTemp} (c)] the periodicity is $\Phi\approx \Phi_0$. Deviations from the periodicity $\Phi/ \Phi_0=1$ can occur because of the small but finite coupling to the discrete levels.

The underlying periodicity of the A-B resonances present in Fig.~\ref{CondTemp} emerges more clearly at higher temperatures. Here, the interference of electron trajectories that do not enclose an integer multiple of the moir\'e unit cell area are averaged out, because they accumulate a different dynamical phase \cite{Virtanen2011,DeBeule2020}.
Although this suppression affects both scattering mechanisms, we observe that for $T/T_l=0.1$ the period $\Phi/\Phi_0=1$ is restored faster for the conductance away from $E_F=20\epsilon$. Whereas close to a resonance $E_F=1.1\epsilon$ or $E_F\approx0.997(\epsilon+U)\epsilon$, the period remains $\Phi/\Phi_0=2$.

\begin{figure}
	\includegraphics[width=0.5\textwidth]{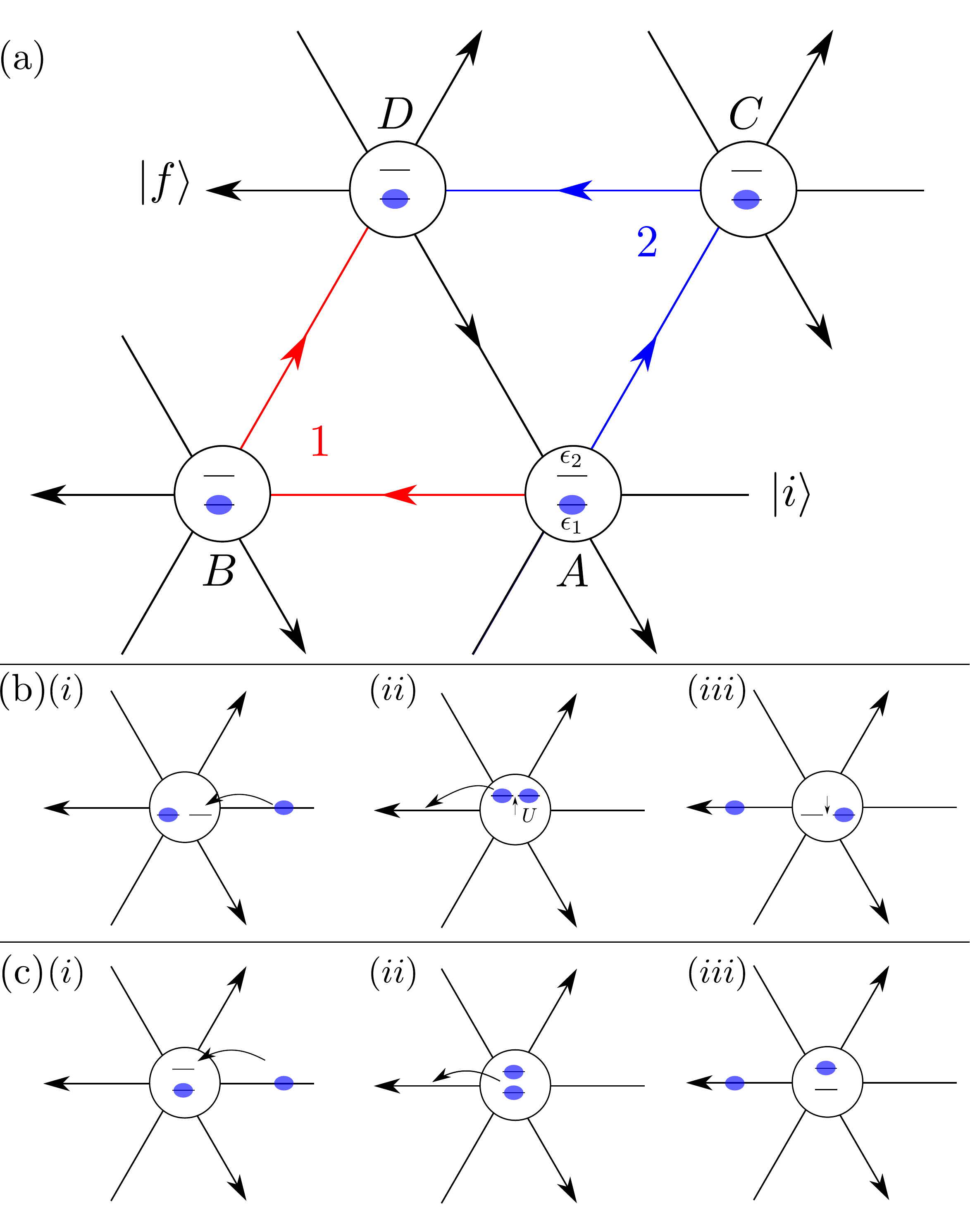}
	\caption{Panel (a): Reduced system with four $AA$ regions $A,B,C,D$ around one moir\'e unit cell. 
 We highlight in blue and red the top and and bottom paths \textcolor{red}{1} and \textcolor{blue}{2}, through which the electron interferes. Panels (b) and (c): Elastic (b) and inelastic (c) processes changing the occupation of the levels.
 } \label{tmatrixsystem}
\end{figure}

{\bf Decoherence effects}--- So far we have assumed that electrons propagate across the network without modifying the occupation on the discrete levels, which is implicitly assumed in the mean-field approximation. However, in the actual system nothing prevents to change the occupation of the discrete levels in a cotunneling process through an $AA$ region, see two examples in Fig.~\ref{tmatrixsystem}(b) and~(c). These changes of occupation events can lead to a ``which path'' decoherence process, which can reduce significantly the Aharonov-Bohm resonances.

To understand the impact of the decoherence processes, we study the smallest network that encloses a moir\'e unit cell containing two paths 1 and 2, see Fig.~\ref{tmatrixsystem} (a).
In this setup, the total probability for going from $\ket{i}$ to $\ket{f}$ is given by 
\begin{equation}
P=\sum_{j=1}^{N_p} \left(P_1^j + P_2^j + 2\mathcal{R}\{\sqrt{P_1^jP_2^j}\braket{\chi_1^j|\chi_2^j}e^{-i \frac{\Phi}{\Phi_0}}\}\right). \label{eq.decoherence}
\end{equation}
The subindex $j$ runs over all possible final configurations of the $AA$ regions ($N_p$), and $\mathcal{R}$ denotes the real part. Here, $P_i^j$ is the probability that an electron moves through the system over the path $i$, whereby the system ends in the configuration $j$. 
The last term in Eq.~\eqref{eq.decoherence} accumulates a Peierls phase due to the magnetic flux piercing the enclosed $AB$ and $BA$ regions and it is responsible for the A-B resonances obtained above. Precisely, this term is susceptible of being canceled by  a change of occupation, which enters via the overlap  $\braket{\chi_1^j|\chi_2^j}$. Here, $\ket{\chi_\alpha^j}$ accounts for the state of the discrete levels after the transition with configuration $j$ of path $\alpha=1,2$. 

The cancellation of the interference term depends on whether it is possible to ``label'' the path the electron has taken via the change in the occupation of the discrete levels.
For example, if the final state of the $AA$ regions exhibits no change in the occupation, the state $\ket{\chi_\alpha}$ then reads 
\begin{equation}
\ket{\chi_\alpha} = d^\dagger_{D1}d^\dagger_{C1}d^\dagger_{B1}d^\dagger_{A1}\ket{}=\ket{i},
\end{equation}
where $d^\dagger_{Li}$ creates an electron in the $AA$ region ${L=A,B,C,D}$ in the energy level $i=1,2$. 
In this case, both of the states $\ket{\chi_\alpha}$ are equal and therefore $|\langle \chi_1 |  \chi_2 \rangle|=1$. 
This also applies if both paths change the occupation at $A$ or $D$. In these cases, there is no ``labeling'' possible on the path.
However, if we change the occupation at $B$ or $C$ for paths 1 and 2 respectively, the interference term cancels because $\braket{\chi_1^j|\chi_2^j}=0$.

We differentiate between elastic and inelastic ``labeling'' events, noting that the latter has a negligible impact because the number of inelastic ``labeling'' events is constrained by the bias voltage. Note that the bias voltage is bounded to the gap opened by the interlayer bias. Thus, for small bias voltage and a large system size, the number of ``labeling'' events is negligible compared to the number of times an electron can encircle a closed path and gather a magnetic flux. In turn, the situation is different for the elastic ``labeling'' events, where the number of processes is not bounded by the bias voltage \cite{Nazarov1990, Nazarov2009Book}. We can see its impact in the simplest network considered in Fig.~\ref{tmatrixsystem}(a), with $N_p=16$ corresponding to all possible final configurations of the $AA$ regions, out of which the number of processes that do not ``label'' the path is only 4. 
Therefore, in this scenario we expect a reduction of the A-B conductance maxima when the Fermi energy is close to  resonance with the localized energy levels, i.e.~$|E_F-\epsilon|\lesssim |\tau|^2E_l$ or $|E_F-\epsilon-U|\lesssim |\tau|^2E_l$.
Note that for $|E_F-\epsilon|\gg |\tau|^2E_l$ and $|E_F-\epsilon-U|\gg |\tau|^2E_l$, the probability of going through the discrete levels is negligible since it is more preferable to deflect to a neighbouring link and therefore, the ``labeling'' events are less likely to take place.

\section{Conclusions} \label{Conclusion}

Inspired by microscopic calculations on minimally twisted bilayer graphene under an interlayer bias\cite{Prada2013,Ramires2018,Walet_2019}, we have modeled phenomenologically a triangular network of valley chiral modes that propagate along the $AB/BA$ interfaces, forming the sides of the triangles, and scatter at the $AA$ regions, placed at the triangles vertices. In contrast to previous phenomenological models\cite{DeBeule2020,DeBeule2021} that consider a negligible density of states at the $AA$ regions, we model them using a discrete density of states with $N$ energy levels, which are coherently coupled to the valley chiral modes. In addition, we have considered an energy-independent deflection process that accounts for the overlap of the neighboring chiral modes wave functions. This overlap becomes finite only in the proximity of the $AA$ regions, where incoming and outgoing valley chiral states approach one another. To combine these two processes in a single $S$ matrix, we have used the generalized Weidenm\"uller formula\cite{Tripathi2016}, yielding an energy-dependent $S$ matrix with two asymptotic limits depending on the energy difference between the position of the discrete levels and the propagating modes: close to resonance, the scattering processes with the discrete levels become relevant, while far from the resonances deflection processes dominate. This approach allows us to include Coulomb interactions on a mean-field level at the $AA$ regions.

We have investigated the impact of the coupling between the discrete states at the $AA$ regions and the valley chiral modes in the energy spectrum of the network and on the magnetoconductance. We have found that close to the resonance, the discrete levels hybridize with the chiral ZZ modes ($\phi\approx0$) or the flat bands ($\phi\approx\pi/2$), giving rise to anticrossings in the energy spectrum and dips ($\phi\approx0$) or peaks ($\phi\approx\pi/2$) in the conductance. Far from resonance, the system is dominated by the deflection processes.

The two limits, close and far from the resonance, can be differentiated not only by the relative size of their conductances, but also by the periodicity of the magnetoconductance, where Aharonov-Bohm resonances arise at multiples $2/n$ of the magnetic flux $\Phi/\Phi_0$ with $n$ being integer, which originate from trajectories encircling an area $n\mathcal{A}/2$. This difference arises due to the presence of chiral ZZ modes in the phenomenological model, which couple mainly to other parallel chiral ZZ modes and therefore cannot encircle a single $AB$ or $BA$ region, such that the periodicity of the magnetoconductance is reduced. In contrast, close to resonance, the periodicity of all multiples are possible, due to the absence of ZZ modes. This difference becomes more visible at higher temperatures. Here, processes that do not enclose an integer multiple of the moir\'e unit cell area accumulate a finite dynamical phase, and therefore, at finite temperatures they become averaged out.

Finally, using qualitative arguments we have shown that when the Fermi energy is close to resonance with the discrete energy levels, elastic cotunneling events can give rise to a change in the occupation of the discrete levels, inducing ``which path'' decoherence processes, suppressing the Aharanov-Bohm resonances present in the conductance.

\subsection*{Acknowledgments}
We thank T. Schmidt for fruitful discussions. We gratefully acknowledge the support of the Braunschweig International Graduate School of Metrology B-IGSM and the DFG Research Training Group 1952 Metrology for Complex Nanosystems. F.D. and P.R. gratefully acknowledge funding by the Deutsche Forschungsgemeinschaft (DFG, German Research Foundation) within the framework of Germany’s Excellence Strategy – EXC-2123 Quantum - Frontiers – 390837967. C.D.B. acknowledges the Luxembourg National Research Fund (FNR) (project No.\ 16515716).

\appendix 
\section{Self-consistent calculation of the occupation number}\label{App.Selfcon}

We calculate the occupation number of the energy levels in the $AA$ region self-consistently. To do so we use the Dyson equation to find the Green's function of these energy levels. We define first the following Green's functions in real space and time domain
\begin{align}
g_{ii}(x,x';t)&= -i \theta(t) \langle\{\psi_{i}(x,t),\psi_i^\dagger(x',0)\}\rangle_0,\\ 
g^j_{DD}(t)&= -i \theta(t) \langle\{d_j(t),d_j^\dagger(0)\}\rangle_0.
\end{align}
The Green's functions $g$ belong to the free system without coupling. So we have an isolated $AA$ region and isolated chiral modes. This is shown by the index $0$ of the expectation value. To calculate these, we use the equation of motion method. The free $AA$ region is described by $H_{D}$ in Eq.~\eqref{eq:HDL} and the free chiral mode by $H_0$ in Eq.~\eqref{eq.H0} in the main text. We find after a Fourier transform from time to energy domain
\begin{align}
&g^j_{DD}(E) =  \frac{1}{E-\epsilon_j-U \langle n_{j'}\rangle}, \label{DotGreen}\\
&g_{ii}(x,x';E)=\frac{-i}{\hbar v_F} \Theta(x-x') e^{i E (x-x')/\hbar v_F}.
\end{align}
Note that the solution for the free chiral mode is the same as for a helical edge state \cite{Rizzo2016,Dominguez2018}.

For the case with finite coupling between $AA$ region and modes, see Eq.~\eqref{VI} in the main text, we define the following Greens functions
\begin{align}
&G_{DD}^j(t)= -i \theta(t) \langle\{d_j(t),d_j^\dagger(0)\}\rangle\\
&G^j_{iD}(x;t)= -i \theta(t) \langle\{\psi_i(x,t),d_j^\dagger(0)\}\rangle.
\end{align}
To calculate these we use the Dyson equation in the following form
\begin{align}
&G^j_{DD}(E) = g^j_{DD}(E) +\sum_{i=1}^{6} g^j_{DD}(E) H_{T,Di}^j G^j_{iD}(x_{AA};E), \label{GDD}\\
&G^j_{iD}(x;E) =g_{ii}(x,x_{AA};E) H_{T,iD}^j G^j_{DD}(E).\label{GID}
\end{align}
We assume that the occupation numbers of the dot levels is a local property of each $AA$ node. Therefore, we have only considered tunneling processes between the dot levels and the chiral channels, see Eq.~\eqref{VI}. Moreover, due to the infinitesimal bias applied, we assume that each chiral channel is in equilibrium with external reservoirs (setting the Fermi energy $E_F$). It also means that deflections between chiral channels ($S_0$) do not contribute as they connect different $AA$ nodes.

Plugging Eq.~\eqref{GID} into Eq.~\eqref{GDD} we find 
\begin{align}
G^j_{DD} &= \frac{g^j_{DD}(E)}{1- g^j_{DD}(E) |t_D|^2\sum_i g_{ii}(x_{AA},x_{AA},E)}\\
&=\frac{g^j_{DD}(E)}{1 + g^j_{DD}(E) \frac{6i|t_D|^2}{\hbar v_F}}.
\end{align}
We have used $H_{T,Di}^jH_{T,iD}^j=|t_D|^2$ like in the main text for all $i$ and $j$.
Therefore, we find the following self energy
\begin{align}
\Sigma = -\frac{6i|t_D|^2}{\hbar v_F} = -i \Gamma/2,
\end{align}
where $\Gamma$ leads to a broadening of the energy levels in the $AA$ region. This holds true for both energy levels.

Note that the parameter $\Gamma$ is calculated assuming that the density of states of the network is constant. As we have shown, the spectrum can exhibit gap openings, resulting in an energy-dependent density of states. Therefore, a more accurate approach would involve the self-consistent calculation of the coupling $\Gamma(E)$. However, due to the perturbative limit of small coupling we are analyzing, we do not expect to see qualitative differences between both approaches, since the coupling parameter $\Gamma$ can only decrease in the self-consistent limit.

We calculate the average occupation self-consistently assuming that the discrete levels are in equilibrium with respect to the external leads that set the Fermi energy $E_F$, namely\cite{Bruus2004a} ($i\neq i'$)
\begin{align}
\langle \hat{n}_i\rangle = \int_{-\infty}^{\infty} \frac{dE}{2\pi} f_0(E)\frac{\Gamma}{(E-\epsilon_i-U\langle \hat{n}_{i'}\rangle)^2+(\Gamma/2)^2 }  \label{selfconsistentT}
\end{align} 
with $f_0(E)$ the Fermi-Dirac distribution 
\begin{equation}
	f_0(E) = \frac{1}{\exp\left[(E-E_F)/k_BT\right]+1}.
\end{equation} 
In the limit of low temperature $k_BT\ll \Gamma$, we find
\begin{align}
\langle \hat{n}_i\rangle &\approx \int_{-\infty}^{E_F} \frac{dE}{2\pi} \frac{\Gamma}{(E-\epsilon_i-U\langle \hat{n}_{i'}\rangle)^2+(\Gamma/2)^2 } \nonumber \\
&= \frac{1}{2}+\frac{1}{\pi} \arctan\left(\frac{E_F-\epsilon_i-U\langle  \hat{n}_{i'}\rangle}{\Gamma/2}\right).\label{selfconsistent}
\end{align}
Here, $\Gamma/2 = 6|t_D|^2/\hbar v_F$ is the broadening produced by the coupling to the chiral modes. 
Both equations can be solved self-consistently by plugging the solution for $\langle \hat{n}_i\rangle$ into the equation of $\langle  \hat{n}_{i'}\rangle $ and vice versa until these values have converged.

\section{Scattering matrices}\label{App.details}

\subsection{Generalized Weidenm\"uller formula}\label{App.detailsMW}

The generalized Weidenm\"uller formula used in Ref.~\onlinecite{Tripathi2016} and in the main text  Eq.~\eqref{eq.MWhere} is equivalent to 
\begin{align}
S(E)=-\mathbb{1}+2\left[\mathbb{1}+i\pi\nu R^t+\frac{i\nu \pi}{2} g(E) (\mathcal{W}^\dagger \mathcal{W})^t\right]^{-1}, \label{eq.Sequiv}
\end{align}
with 
\begin{align}
g(E)=\frac{1}{E-\epsilon_1-U\langle \hat{n}_2\rangle } +\frac{1}{E-\epsilon_2-U\langle \hat{n}_1\rangle } . \label{eq.fe}
\end{align}
Using this simplified expression it is easier to estimate the regimes in which either the deflection process or the scattering via the discrete energy levels dominates. For the inverse of the sum of two matrices $A$ and $B$, where $A$ and $A+B$ are invertible and $B$ has rank 1, one can write\cite{Miller1981}
\begin{equation}
	(A+B)^{-1}= A^{-1} - \frac{1}{1+q}A^{-1}BA^{-1},
\end{equation}
where $q=\text{Tr}\left(BA^{-1}\right)\neq-1$. For equal couplings the matrix $B=\frac{i\nu \pi}{2} g(E)(\mathcal{W}^\dagger \mathcal{W})^t$ has rank 1, because all entries are the same. For $R=0$ we can write $A=\mathds{1}$ and therefore
\begin{equation}
	S(E)= \mathds{1}-\frac{i\pi \nu g(E)}{1+6i\pi\nu g(E)|t_D|^2}(\mathcal{W}^\dagger \mathcal{W})^t.
\end{equation}

For finite $R\neq0$, we can write $A= \mathds{1}+i \pi \nu R^t$ and find 
\begin{equation}
	q= \text{Tr}\left(BA^{-1}\right)= 3i\pi \nu |t_D|^2 g(E). \label{eq.g}
\end{equation}
Note that we are including $\phi$ in the propagation along the $AB/BA$ interfaces and thus, Eq.~\eqref{eq.g} does not depend on $\phi$.
Far from the resonance $g(E)\rightarrow0$, we find $B\rightarrow0$, $q\rightarrow0$ and therefore
\begin{equation}
	S(E)= -\mathds{1}+2 \left(\mathds{1}+i \pi \nu R^t\right)^{-1}= S_0.
\end{equation}
This equation is equivalent to Eq.~\eqref{eq.S0}. This proves, that far from the resonance the scattering matrix is dominated by the deflection processes. Close to the resonance, however, we have a scattering matrix that contains both parts, the deflection and the scattering with the energy levels. The scattering matrix exactly at resonance $g(E)\rightarrow\infty$ is given in App.~\ref{App.res}. 

We want to understand in which energy window the coupling to the energy levels becomes relevant in the scattering matrix. Using Eq.~\eqref{eq.Sequiv} and \eqref{eq.g}, we find, that we are close to resonance, if $g(E)|t_D|^2\ll \hbar v_F$ or in other words
\begin{equation}
	\Delta E\sim x\ll |t_D|^2/\hbar v_F,
 \end{equation} 
 where $x=1/g(E)$ and $\Delta E=E-\epsilon$ or $\Delta E=E-\epsilon-U$.

\subsection{$S_0$ and its relation to $R$}\label{App.detailsTmatrix}

As we have seen, $S_0$ can be expressed as the product $S_0 = \left(1+i\pi \nu R\right)^{-1}\left(1-i\pi \nu R\right)$. Here, we give the $R$ matrix as a function of the parameters entering in the phenomenological $S$ matrix

\begin{align}
		&i \pi \nu R = i\frac{1}{2 \sqrt{P_{d1}}\sin(\phi)}\begin{pmatrix}
		t_f & t_l & t_r\\
		t_r & t_f & t_l\\
		t_l & t_r & t_f
		\end{pmatrix},\\
		&t_f = \begin{pmatrix}
		1+2 \sqrt{P_{d1}}\cos(\phi) & 
		  \sqrt{P_{d21}}-\sqrt{P_{d22}}\\
		 \sqrt{P_{d21}}-\sqrt{P_{d22}} &
		1-2 \sqrt{P_{d1}}\cos(\phi)
		\end{pmatrix},\\
		&t_l=\begin{pmatrix}
		-1 &  \sqrt{P_{d21}}+\sqrt{P_{d22}}\\
		 -\left(\sqrt{P_{d21}}+\sqrt{P_{d22}}\right)& 
		-1\\
		\end{pmatrix}
\end{align}
with $t_l=t_r^t$.

\subsection{The scattering matrix without deflection} \label{App.details1p1}
Using the Weidenm\"uller formula Eq.~\eqref{eq:SD} $(R=0)$ results in a $S$ matrix that can be written as
\begin{align}
	S_D&=\begin{pmatrix}
	s_f & s_d & s_d\\
	s_d & s_f & s_d\\
	s_d & s_d & s_f\\
	\end{pmatrix},
\end{align}
with
\begin{align}
	s_f=& \left(1-\frac{2i\pi \nu g(E) |t_D|^2}{1+6i\pi\nu g(E) |t_D|^2}\right)\sigma_0\\
	&-\frac{2i\pi\nu g(E)|t_D|^2}{1 + 6i\pi \nu g(E)|t_D|^2}\sigma_x,\\ s_d=& -\frac{2i\pi\nu g(E)|t_D|^2}{1 + 6i\pi\nu g(E)|t_D|^2}\left(\sigma_0+\sigma_x\right),
\end{align}
where $\sigma_i$ are Pauli matrices. With the unitary transformation $U= \exp(-i \pi/4 \sigma_y)\otimes \mathds{1}_3$ we find that
\begin{align}
	\tilde{S}_D&= \begin{pmatrix}
	S_1 & 0\\0 & S_2\\	\end{pmatrix}
\end{align}
where $S_1$ and $S_2$ are $3 \times 3$ matrices. $S_2$ has only values on the main diagonal equal to 1 and $S_2$ is of the form of the Efimkin-MacDonald model \cite{Efimkin2018,DeBeule2021Floquet}. This proves that this scattering matrix describes a combination of three free chiral modes and a single channel Efimkin-MacDonald model.

\subsection{Scattering matrix at resonance} \label{App.res}
\begin{figure}
	\includegraphics[width=0.4\textwidth]{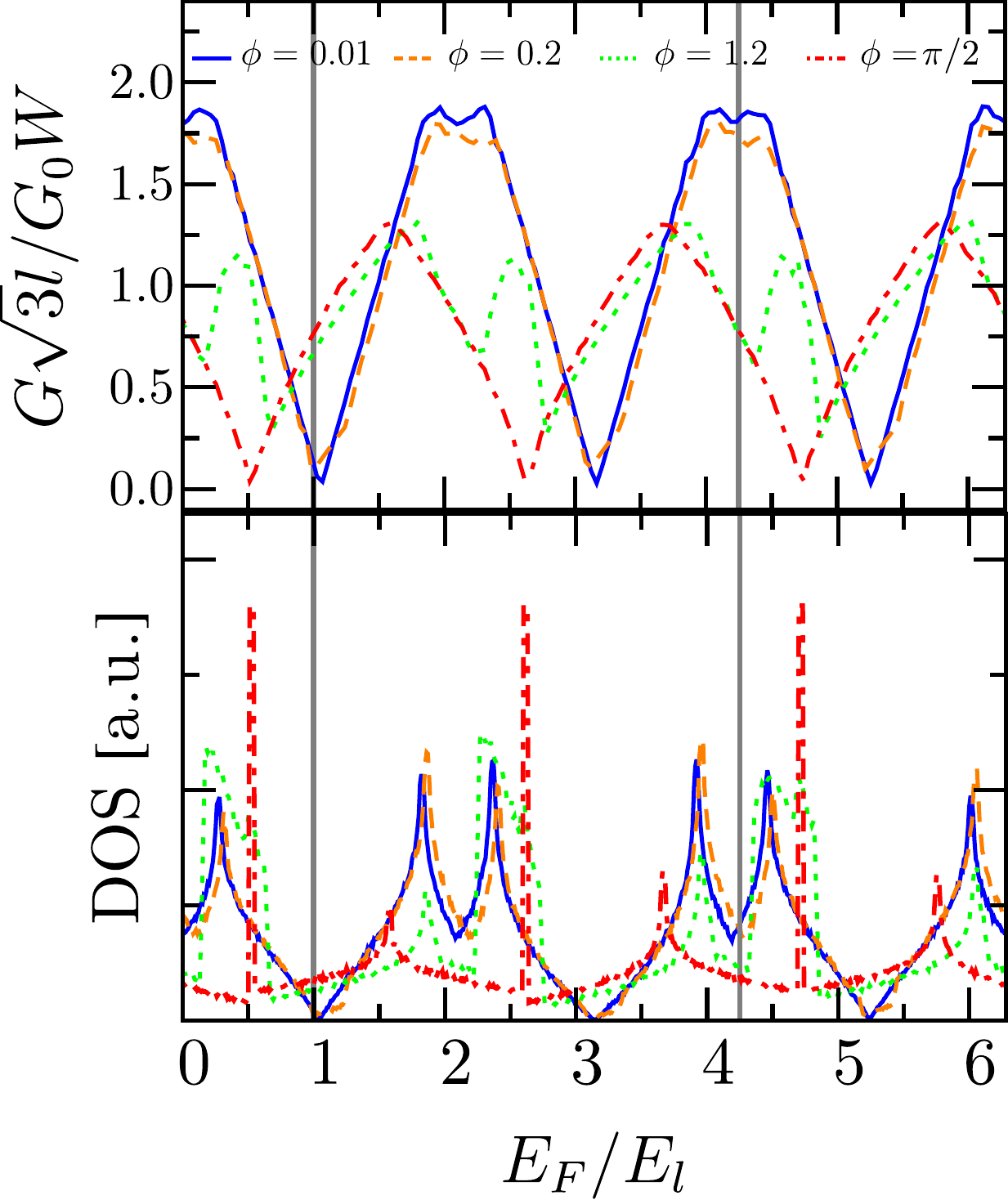}
	\caption{Top: Conductance of a network strip at resonance with the discrete levels, i.e. for $E_F=\epsilon$ or $E_F=\epsilon+U$ and different values of $\phi$. We see depending on the energy a higher or lower conductance. The values used in the main text of $\epsilon=\hbar v_F/l$ and $\epsilon+U=4.25\hbar v_F/l$  are marked with vertical grey lines. Bottom: Density of states of the same network. Peaks in the density of states correspond to saddle points or completely flat bands in the band structure. The latter do not contribute to the conductance.} \label{fig.Condres}
\end{figure}

We can rewrite the scattering matrix from Eq.~\eqref{eq.MWhere} at resonance $E_F=\epsilon$ or $E_F=\epsilon+U$  in the following way:
	\begin{align}
S= \begin{pmatrix}
s_f & s_d & s_d^t\\
s_d^t & s_f & s_d\\
s_d & s_d^t & s_f
\end{pmatrix} \label{eq.Sres}
\end{align}
with
\begin{align}
s_f&= 
\begin{pmatrix}
-\frac{2}{3} & 0 \\
0 & 0 \\
\end{pmatrix},\\
s_d & =
\begin{pmatrix}
-\frac{1}{6} & -\frac{1}{2} \\
\frac{1}{2} & -\frac{1}{2}  \\
\end{pmatrix}.
\end{align}
We plot in Fig.~\ref{fig.Condres} the conductance of a network strip, where the scattering matrix for all nodes in the strip is the one at resonance, given in Eq.~\eqref{eq.Sres}. That means we calculate the conductance at resonance. The energy dependence is entering due to the dynamical phase accumulated along the links. We can see for different values of $\phi$, which is accumulated along the links, a different behavior. We see a periodic dip and peak structure. This is in agreement with the density of states of such a system. Big spikes in the density of states correspond to flat bands, which are not contributing to the conductance due to the lack of velocity of electrons in these states.

We have chosen for our calculations one resonance point inside of a conductance dip $\epsilon=\hbar v_F/l$ and one at a conductance peak $\epsilon+U=4.25\hbar v_F/l$.

\section{Transport calculations} \label{App.Transport}
Here, we explain how we have calculated the transmission function $\mathcal{T}(E)$ for our transport calculations in Eq.~\eqref{eq.Cond} by the combination of scattering matrices. We show the general idea how to combine the first two scattering matrices. Finally, we explain from this the recursive loop that we have implemented for our calculations.
\begin{figure}
	\includegraphics[width=0.4\textwidth]{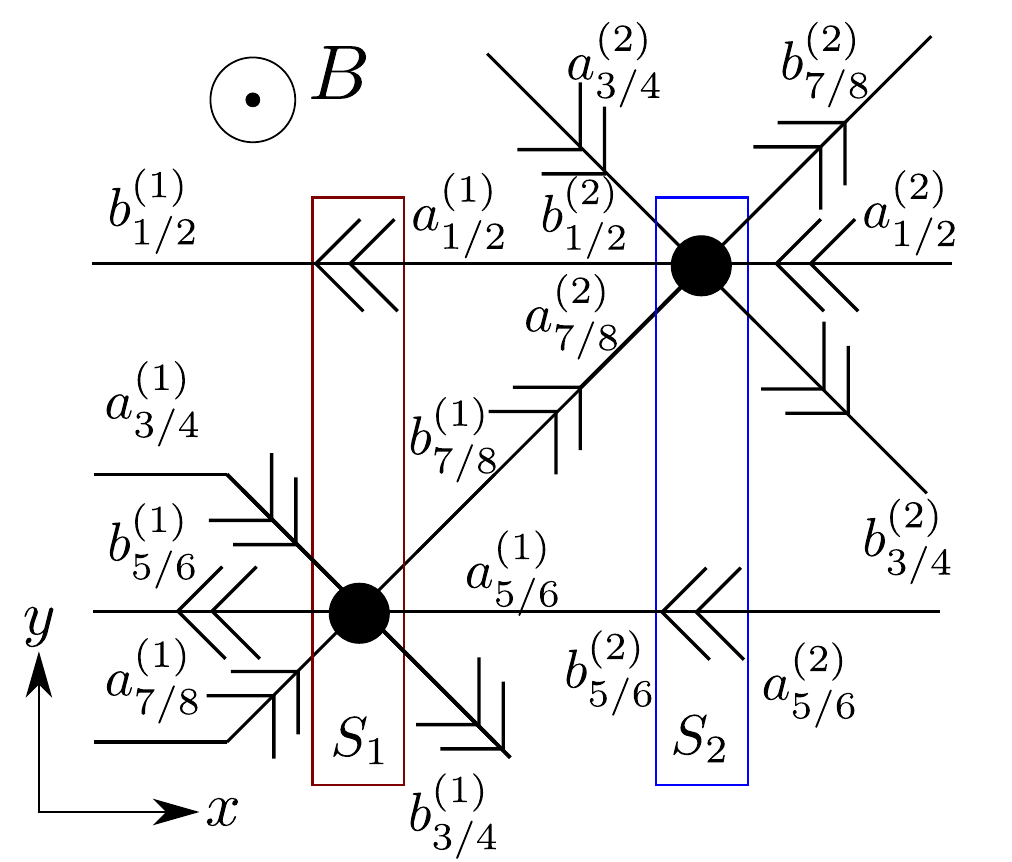} \caption{Part of the network in Fig.~\ref{fig.Networkmodel} with the basis elements for the first two scattering matrices. $a(b)_{x/y}^{(i)}$ are the incoming (outgoing) modes of the scattering matrix $S_i$.} \label{fig.NetworkTransport}
\end{figure}

The first scattering matrix $S_1$ (see Fig.~\ref{fig.Networkmodel} and Fig.~\ref{fig.NetworkTransport}) is given by
\begin{equation}
S_1=\begin{pmatrix}
\mathds{1} & 0\\
0 & S(E)
\end{pmatrix},\label{eq.S1}
\end{equation}
where $S(E)$ is given in Eq.~\eqref{eq.MWhere}. The used basis $(b_{1/2}^{(1)},b_{3/4}^{(1)},b_{5/6}^{(1)},b_{7/8}^{(1)})^t=S_1 (a_{1/2}^{(1)},a_{3/4}^{(1)},a_{5/6}^{(1)},a_{7/8}^{(1)})^t$ is depicted in Fig.~\ref{fig.NetworkTransport}. Note, that the second scattering matrix $S_2$ can be calculated from $S_1$ by exchanging the first and the third two modes, i.e.
\begin{equation}
S_2=R_{13} S_1 R_{13}, \, R_{13}= \begin{pmatrix}
0 & 0 & \mathds{1} & 0\\
0 & \mathds{1} & 0 & 0\\
\mathds{1} & 0 & 0 & 0\\
0 & 0 & 0 & \mathds{1}\\
\end{pmatrix}.
\end{equation}
With this it connects $(b_{1/2}^{(2)},b_{3/4}^{(2)},b_{5/6}^{(2)},b_{7/8}^{(2)})^t=S_2 (a_{1/2}^{(2)},a_{3/4}^{(2)},a_{5/6}^{(2)},a_{7/8}^{(2)})^t$ like it is depicted in Fig.~\ref{fig.NetworkTransport}.
Now we need to combine these two matrices. Before we look into that we change the basis a little bit to make the process of combining easier. We rewrite $S_1$, so that it fulfills
\begin{align}
\begin{pmatrix}
b_L^{{(1)}}\\
b_R^{{(1)}}\\
\end{pmatrix} = \begin{pmatrix}
r_L^{{(1)}} & t_{LR}^{{(1)}}\\
t_{RL}^{{(1)}} & r_R^{{(1)}}
\end{pmatrix} \begin{pmatrix}
a_L^{(1)}\\
a_R^{(1)}\\
\end{pmatrix}.  \label{eq.S1n}
\end{align}
To bring these notations together we define 
\begin{align}
b_L^{(1)}&:= \begin{pmatrix}
b_{1,2}^{(1)}, b_{5,6}^{(1)}
\end{pmatrix}^t, \,
b_R^{(1)}:= \begin{pmatrix}
b_{3,4}^{(1)}, b_{7,8}^{(1)}
\end{pmatrix}^t,\\
a_L^{(1)}&:= \begin{pmatrix}
a_{7,8}^{(1)}, a_{3,4}^{(1)}
\end{pmatrix}^t,\,
a_R^{(1)}:= \begin{pmatrix}
a_{1,2}^{(1)}, a_{5,6}^{(1)}
\end{pmatrix}^t.
\end{align}
With that we bring the scattering matrix into a block structure with the submatrices $r_{i}$, that contains the reflection processes to the direction $i=L,R$, and $t_{ij}$, that contains the transmission processes from $j=L,R$ to $i=L,R$, where $L$ means left and $R$ means right. Note that $r_{i}$ and $t_{ij}$ are $4\times 4$ matrices. Due to the $C_3$ symmetry of the system, we can write the reflection matrix $r_i$ and the transmission matrix $t_{ij}$ in terms of three submatrices $s_f$, $s_r$ and $s_l$ as
\begin{align}
r_L^{(1)}&:= \begin{pmatrix}
0 & 0\\
s_l & s_r
\end{pmatrix},\,
r_R^{(1)} := \begin{pmatrix}
0 & s_l\\
0 & s_r
\end{pmatrix},\\
t_{RL}^{(1)}&:= \begin{pmatrix}
s_r & s_f\\
s_f & s_l
\end{pmatrix},\,
t_{LR}^{(1)} := \begin{pmatrix}
1 & 0\\
0 & s_f
\end{pmatrix},\\
	s_f &= \begin{pmatrix}
	s_{11} & s_{12}\\
	s_{21} & s_{22}
	\end{pmatrix},\, 
	s_l = \begin{pmatrix}
	s_{13} & s_{14}\\
	s_{23} & s_{24}
	\end{pmatrix}=s_r^t,
\end{align}
where $s_{ij}$ is the matrix element of $S_1$ in Eq.~\eqref{eq.S1}. The second scattering matrix can be written in a similar way as
\begin{align}
\begin{pmatrix}
b_L^{(2)}\\
b_R^{(2)}\\
\end{pmatrix} = \begin{pmatrix}
r_L^{(2)} & t_{LR}^{(2)}\\
t_{RL}^{(2)} & r_R^{(2)}
\end{pmatrix} \begin{pmatrix}
a_L^{(2)}\\
a_R^{(2)}\\
\end{pmatrix}, \label{eq.S2n}
\end{align}
with
\begin{align}
b_L^{(2)}&:= \begin{pmatrix}
b_{1,2}^{(2)}, b_{5,6}^{(2)}
\end{pmatrix}^t,\,
b_R^{(2)}:= \begin{pmatrix}
b_{7,8}^{(2)}, b_{3,4}^{(2)}
\end{pmatrix}^t,\\
a_L^{(2)}&:= \begin{pmatrix}
a_{3,4}^{(2)}, a_{7,8}^{(2)}
\end{pmatrix}^t,\,
a_R^{(2)}:= \begin{pmatrix}
a_{1,2}^{(2)}, a_{5,6}^{(2)}
\end{pmatrix}^t,\\
r_L^{(2)}&:= \begin{pmatrix}
e^{-ik\sqrt{3}l}s_r& s_l\\
0 & 0
\end{pmatrix},\,
r_R^{(2)} := \begin{pmatrix}
e^{ik \sqrt{3}l}s_r & 0\\
s_l & 0
\end{pmatrix},\\
t_{RL}^{(2)}&:= \begin{pmatrix}
s_l & e^{i k \sqrt{3}l}s_f\\
e^{-i k \sqrt{3}l}s_f & s_r
\end{pmatrix},\,
t_{LR}^{(2)} := \begin{pmatrix}
s_f & 0\\
0 & 1
\end{pmatrix}.
\end{align}
We have added in the second scattering matrix also the transversal momentum\cite{DeBeule2020} $0\leq k<2\pi/\sqrt{3}l$. Due to the translational symmetry in $y$ direction, the modes, that leave and enter Fig.~\ref{fig.NetworkTransport} in  $y$ direction, are related by Bloch's theorem. We integrate over the transversal momentum at the end.
Now we need to know how the incoming and outgoing modes are related. We can write
\begin{align}
&a_R^{(1)} = \alpha b_L^{(2)}, \, a_L^{(2)} = \beta_1 b_R^{(1)}, \label{eq.conn}\\
&\alpha = e^{i \frac{E l}{2\hbar v_F}}\left(\mathds{1}_2\otimes e^{i\phi/2\sigma_z}\right),\\
&\beta_n =e^{i \frac{E l}{\hbar v_F}}(\mathds{1}_2\otimes e^{i\phi\sigma_z}) \label{eq.Peierls}\\
&(\exp[(-1)^{n+1}(2n-1) i\pi \Phi/2 \Phi_0\sigma_z]\otimes\mathds{1}_{2}). \nonumber
\end{align}
The matrix $\beta_n$ contains the phase $\phi$ due to the momentum difference, the dynamical phase and the Peierls phase due to the magnetic field of a mode traversing from one node to the next. The matrix $\alpha$ contains the phase $\phi$ and the dynamical phase of a mode traversing in $x$ direction. Therefore it does not accumulate a Peierls phase due to the used gauge $\mathbf{A}=Bx \mathbf{e}_y$. The parameter $n\in \mathds{N}$ will be counted up for every combining step.

With that we can eliminate $b_R^{(1)}$ and $b_L^{(2)}$ in Eq.~\eqref{eq.S1n} and Eq.~\eqref{eq.S2n} with Eq.~\eqref{eq.conn}. We find
\begin{align}
b_R^{(2)}&= t_{RL}^{(2)} \beta_1 Q_1 t_{RL}^{(1)}a_L^{(1)}+(r_R^{(2)}+t_{RL}^{(2)}\beta_1 Q_1 r_{R}^{(1)}\alpha t_{LR}^{(2)})a_R^{(2)},\label{eq.mcomb1}\\
b_L^{(1)}&= (r_L^{(1)}+t_{LR}^{(1)}\alpha Q_2 r_L^{(2)}\beta_1 t_{RL}^{(1)})a_L^{(1)}+t_{LR}^{(1)}\alpha Q_2 t_{LR}^{(2)}a_R^{(2)},\\
Q_{1} &= \left[1- r_R^{(1)}\alpha r_L^{(2)}\beta_1\right]^{-1}, \, Q_{2} =\left[1- r_L^{(2)}\beta_1 r_R^{(1)}\alpha\right]^{-1}. \label{eq.mcomb2}
\end{align}
Using Eq.~\eqref{eq.mcomb1} and \eqref{eq.mcomb2} we write the new $S$ matrix as
\begin{align}
\begin{pmatrix}
b_L^{{(1)}}\\
b_R^{{(2)}}\\
\end{pmatrix} = \begin{pmatrix}
r_L & t_{LR}\\
t_{RL} & r_R
\end{pmatrix} \begin{pmatrix}
a_L^{(1)}\\
a_R^{(2)}\\
\end{pmatrix}.  \label{eq.Combmatrix}
\end{align}
with
\begin{align}
&r_L = r_L^{(1)}+t_{LR}^{(1)}\alpha Q_2 r_L^{(2)}\beta_1 t_{RL}^{(1)}, \label{eq.comb1}\\
&r_R = r_R^{(2)}+t_{RL}^{(2)}\beta_1 Q_1 r_{R}^{(1)}\alpha t_{LR}^{(2)}, \label{eq.comb2}\\
&t_{LR} = t_{LR}^{(1)}\alpha Q_2 t_{LR}^{(2)}, \label{eq.comb3}\\
&t_{RL} = t_{RL}^{(2)} \beta_1 Q_1 t_{RL}^{(1)}. \label{eq.comb4}
\end{align}

With that we have everything to define our general procedure: First we define the scattering matrix and the previous mentioned submatrices. Then we set in the first step
\begin{align}
r_L = r_L^{(1)},\,
r_R = r_R^{(1)},\,
t_{LR} = t_{LR}^{(1)},\,
t_{RL}= t_{RL}^{(1)}, \label{eq.start}
\end{align}
which defines the first slice of the network with the first scattering matrix, see Eq.~\eqref{eq.S1n}.
In the next step we calculate with the initial conditions in Eq.~\eqref{eq.start} the submatrices of the combined scattering matrix of the first and second slice by

\begin{align}
&r_{L,n} =r_L+t_{LR}\alpha Q^{(2)}_{n,b} r_L^{(2)}\beta_1 t_{RL},\label{eq.loop1}\\
&r_{R,n} = r_R^{(2)}+t_{RL}^{(2)}\beta_1 Q^{(2)}_{n,a} r_{R}\alpha t_{LR}^{(2)},\\
&t_{LR,n} =  t_{LR}\alpha Q^{(2)}_{n,b} t_{LR}^{(2)},\\
&t_{RL,n} =  t_{RL}^{(2)} \beta_1 Q^{(2)}_{n,a} t_{RL},
\end{align}
with
\begin{align}
	Q^{(x)}_{n,a}&=\left[1- r_R\alpha r_L^{(x)}\beta_n\right]^{-1},\\
	Q^{(x)}_{n,b}&=\left[1- r_L^{(x)}\beta_n r_R\alpha\right]^{-1}, \, x=1,2.
\end{align}
The index $n$ counts the step in the loop and also where we are in the network, which is important for the Peierls phase, see Eq.~\eqref{eq.Peierls}. In the first step it will be $n=1$.
The third scattering matrix is the same as the first. Therefore the combining procedure is similar. We exchange the matrices with the upper index $(2)$ with $(1)$ and use the already combined part for the matrices with upper index $(1)$ in Eq.~\eqref{eq.comb1}-\eqref{eq.comb4}. That means
\begin{align}
&r_{L,n+1} = r_L^{(n)}+t_{LR}^{(n)}\alpha Q^{(1)}_{n+1,b} r_L^{(1)}\beta_2 t_{RL}^{(n)},\\
&r_{R,n+1} = r_R^{(1)}+t_{RL}^{(1)}\beta_2 Q^{(1)}_{n+1a} r_{R}^{(n)}\alpha t_{LR}^{(1)},\\
&t_{LR,n+1} =t_{LR}^{(n)}\alpha Q^{(1)}_{n+1,b} t_{LR}^{(2)},\\
&t_{RL,n+1} = t_{RL}^{(2)} \beta_2 Q^{(1)}_{n+1,a} t_{RL}^{(n)}.\label{eq.loop2}
\end{align}
With that we have implemented the combining procedure for a system with size $L=1l$. To enlarge it further we repeat this process. That means we set 
\begin{align}
    &r_L=r_{L,n+1}, \,  r_R=r_{R,n+1},\\  &t_{LR}=t_{LR,n+1}, \,  t_{RL}=t_{RL,n+1}
\end{align}
and repeat Eq.~\eqref{eq.loop1}-\eqref{eq.loop2} with $n\rightarrow n+2$ and so on until we reach the desired length of the network strip.

At the end we can calculate the transmission function per unit cell by
\begin{equation}
\mathcal{T}(E) = (\sqrt{3}l/2\pi)\int_{0}^{2\pi/\sqrt{3}l} dk \text{Tr}[t_{RL}^\dagger t_{RL}].
\end{equation}
Note that the transmission function for the whole system and therefore the conductance is the same for both valleys in the limit of $W\gg L$, where $W$ is the width and $L$ is the length of our system \cite{DeBeule2020}.

\end{document}